\documentclass{osa-article}

\journal{osajournal}

\usepackage{amsmath}
\usepackage{textcomp}
\usepackage{url}

\begin{document}

\title{Improved design and experimental demonstration of ultrahigh-Q C${}_\text{6}$-symmetric H1 hexapole photonic crystal nanocavities}

\author{Kenta Takata\authormark{1,2,4}, Eiichi Kuramochi\authormark{1,2}, Akihiko Shinya\authormark{1,2} and Masaya Notomi\authormark{1,2,3,5}}

\address{\authormark{1}Nanophotonics Center, NTT Corporation, 3-1 Morinosato-Wakamiya, Atsugi, Kanagawa 243-0198, Japan\\
\authormark{2}NTT Basic Research Laboratories, NTT Corporation, 3-1 Morinosato-Wakamiya, Atsugi, Kanagawa 243-0198, Japan\\
\authormark{3}Department of Physics, Tokyo Institute of Technology, 2-12-1 Ookayama, Meguro-ku, Tokyo 152-8551, Japan\\
\authormark{4}kenta.takata.ke@hco.ntt.co.jp\\
\authormark{5}masaya.notomi.mn@hco.ntt.co.jp}




\begin{abstract}
An H1 photonic crystal nanocavity is based on a single point defect and has eigenmodes with a variety of symmetric features. Thus, it is a promising building block for photonic tight-binding lattice systems that can be used in studies on condensed matter, non-Hermitian and topological physics. However, improving its radiative quality ($Q$) factor has been considered challenging. Here, we report the design of a hexapole mode of an H1 nanocavity with a $Q$ factor exceeding $10^8$. We achieved such extremely high-$Q$ conditions by designing only four structural modulation parameters thanks to the ${\rm C_{6}}$ symmetry of the mode, despite the need of more complicated optimizations for many other nanocavities. The fabricated silicon photonic crystal nanocavities exhibited a systematic change in their resonant wavelengths depending on the spatial shift of the air holes in units of 1 nm. Out of 26 such samples, we found eight cavities with loaded $Q$ factors over one million ($1.2 \times 10^6$ maximum). We examined the difference between the theoretical and experimental performances by conducting a simulation of systems with input and output waveguides and with randomly distributed radii of air holes. Automated optimization using the same design parameters further increased the theoretical $Q$ factor by up to $4.5 \times 10^8$, which is two orders of magnitude higher than in the previous studies. Our work elevates the performance of the H1 nanocavity to the ultrahigh-$Q$ level and paves the way for its large-scale arrays with unconventional functionalities.
\end{abstract}

\section{Introduction}
Photonic crystal nanocavities (PCNs) in dielectric slabs are a particular series of optical resonators that exhibit both strong light confinement and small modal volumes \cite{PhC2008,Painter1999,Srinivasan2002,Akahane2003,Notomi2004,Yoshie2004,Song2005,Englund2005,Kuramochi2006,Takahashi2007,Kuramochi2008,Notomi2008}. These features enable intense light-matter interactions, which make PCNs very useful for extremely low-power photonics \cite{Matsuo2010,Takeda2013,Shakoor2014}, on-chip nonlinear optics \cite{Notomi2005,Matsuda2011,Takahashi2013} and quantum optics\cite{Englund2005_2,Nomura2010,Liu2018}. Integration of PCNs also opens a route to functional nanophotonic devices, such as slow light waveguides \cite{Yariv1999,Notomi2008_2,Kuramochi2018}, and all-optical switches \cite{Tanabe2005,Nozaki2010,Nozaki2013}, memories \cite{Tanabe2007,Nozaki2012,Kuramochi2014}, and transistors \cite{Nozaki2019}, which are potential for information processing.

An H1 PCN comprises a vacancy of a single lattice element \cite{Ryu2003,Kim2004,Tanabe2007_2,Takagi2012}. Such a point defect structure takes over the spatial symmetry of its host system. Thus, the eigenmodes of the Maxwell equations for the H1 nanocavity are also those for the symmetry operations in the entire point group of the photonic crystal (PhC) \cite{Sakoda2005}. As a result, they are analogous to atomic orbitals in terms of their symmetric properties, and hence, coupled H1 PCNs work as good photonic emulators of molecules and tight-binding lattices including basis functions \cite{Yariv1999,Altug2004}. Because their evanescent couplings, resonant frequencies and radiation losses can be controlled by structural modulation, PCNs can also be combined with unconventional functionalities emerging in non-Hermitian and topological physics \cite{Takata2017,Takata2018,Han2019,Duggan2020,Takata2021,Fong2021,Takata2022,Hentinger2022,Ozdemir2019,Ota2020}. In particular, arrays of H1 PCNs may pave the way for large-scale two-dimensional crystalline systems \cite{Szameit2011,Kremer2019,Wu2015,Noh2018,Li2020,Khanikaev2017}. This potential is in stark contrast to most other PCNs based on linear defects, which are less symmetric and thus limited in their coupling profiles.

However, it is generally more difficult for a smaller PCN to have an ultrahigh $Q$ factor. Narrower field distributions in real space result in broader ones in reciprocal space. Parts of such modes tend to reside in the light cone (LC) and hence turns into radiation fields, namely losses \cite{Srinivasan2002}. We showed two decades ago that a hexapole mode of the H1 nanocavity in a triangular-lattice PhC slab could have a theoretical $Q$ factor up to $3 \times 10^6$, unlike the other eigenmodes \cite{Ryu2003,Kim2004}. However, this record was not broken even with an algorithmic optimization \cite{Minkov2014}. Moreover, the experimental counterpart was an order of magnitude smaller, namely $3 \times 10^5$ \cite{Tanabe2007_2}. Unfortunately, there values compared disadvantageously to those of PCNs with larger defect regions \cite{Taguchi2011,Lai2014,Simbula2017,Benevides2017,Ashida2018}. The lack of tightest light confinement seems to be a significant obstacle to using large-scale H1 nanocavity arrays, for example, to enhance light-matter interactions with bulky coupled modes, and to make robust optical circuits with topological edge states.

In this article, we design, analyze and experimentally examine the hexapole mode of an H1 PCN with a theoretical $Q$ factor ($Q_{\rm th}$) over $10^8$, on the basis of our latest prototype for studying non-Hermitian physics \cite{Takata2022}. Structural modulation in the design maintains the ${\rm C_{6v}}$ symmetry of the PCN, which the hexapole mode also respects. As a result, we find that we can dramatically increase the $Q$ factor just with four optimization parameters. By elaborating the dependence of $Q_{\rm th}$ on major three parameters in a simulation, we clarify that such extremely high-$Q$ conditions form a region with some width in the parameter space. Here, we obtained a hexapole mode with $Q_{\rm th} = 1.4 \times 10^8$ and a modal volume ($V$) of $0.72 (\lambda/n)^3$. We also compare its field profiles with those of another H1 PCN based on a previous study in real and reciprocal spaces. 

We experimentally investigated a series of silicon (Si) H1 PCNs with different spatial shifts of air holes. These samples exhibited a systematic variation in their resonant wavelengths, indicating that undesired variations in the positions of air holes were restricted. We found that eight such PCNs out of 26 had loaded $Q$ factors ($Q_{\rm exp}$), which include the effects of the input and output waveguides, of over one million. The best sample had $Q_{\rm exp} = 1.2 \times 10^6$, and the cavity's intrinsic $Q$ factor ($Q_{\rm i}$) was estimated to be $Q_{\rm i} = 1.5 \times 10^6$. We also performed a simulation of the system with randomly varying radii of the air holes to close the gap between $Q_{\rm th}$ and $Q_{\rm exp}$.

Finally, we performed an automated optimization to further improve $Q_{\rm th}$. Here, we added the hole radius of the background PhC as a parameter and found $Q_{\rm th} = 4.5 \times 10^8$, which is more than a hundred times those in the previous design. Our results show that the highly symmetric hexapole mode can achieve both an extremely high $Q_{\rm th}$ and a very small $V$ with an inexpensive optimization. It enables ultrahigh $Q_{\rm exp}$ ($> 10^6$) of H1 PCNs and will open up their various applications. 

The remainder of this paper is organized as follows. Section \ref{sec:design} shows the design and modal properties of our H1 PCN. Section \ref{sec:experiment} presents experimental results, and numerically analyzes and discusses them. The automated optimization and resultant impact on the hexapole mode are summarized in Sec. \ref{sec:optimization}. Section \ref{sec:discussion} discusses fundamental limitations on the $Q$ factors of nanocavities, including ours. Section \ref{sec:conclusion} concludes this study.
\begin{figure}[h!] 
	\centering\includegraphics[width=11.5cm]{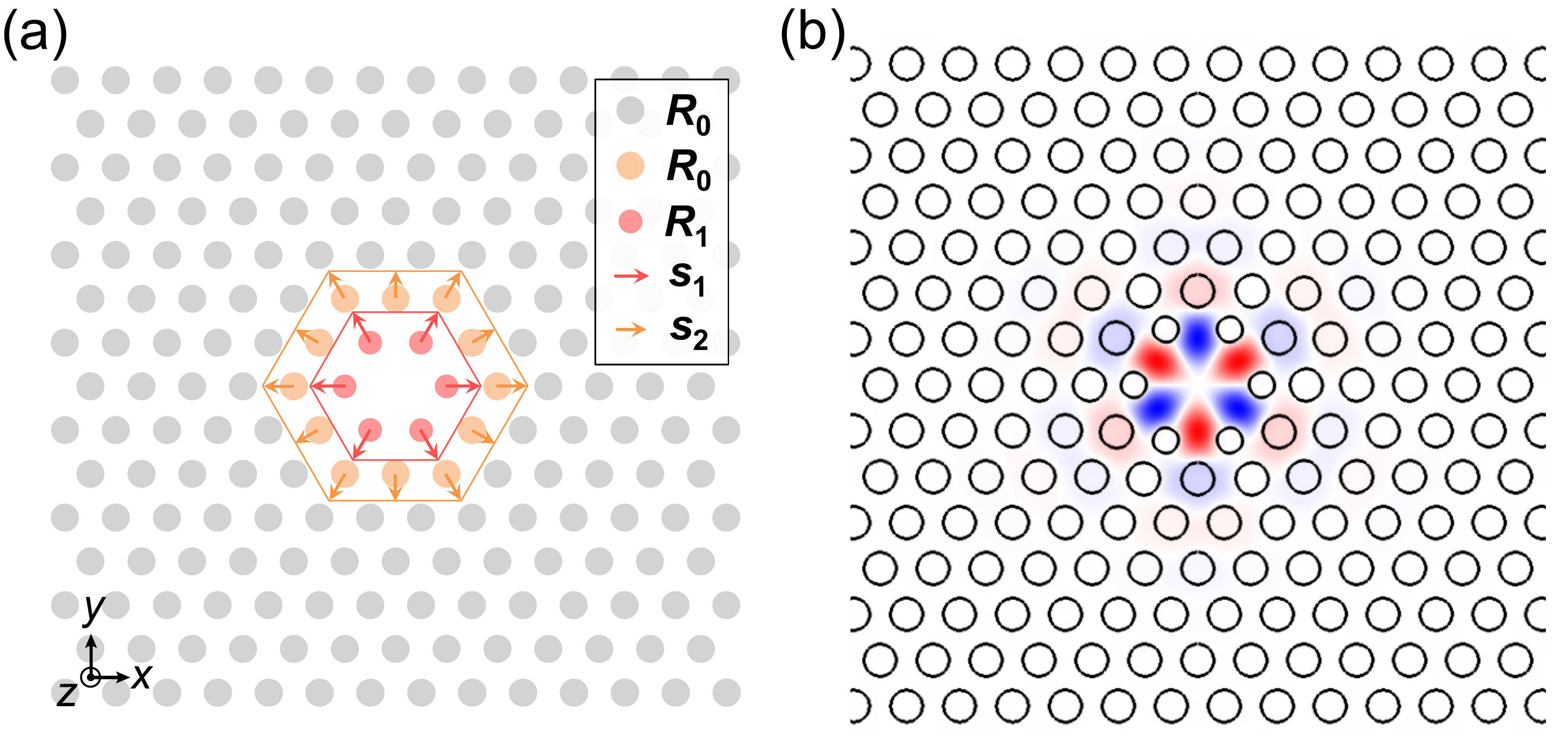}
	\caption{(a) Design of H1 PCN based on structural modulation of the innermost and second innermost layers of air holes with reference to the single point defect (colored red and orange, respectively). $R_0$ is the radius of the holes for the background PhC and the second layer, and $R_1$ that for the innermost holes. $s_1$ is a radial shift of the innermost layer directed outward from the lattice points, and $s_2$ is that for the second innermost layer with its regular hexagonal alignment kept. (b) $H_z$ field distribution of hexapole mode.} \label{fig:design}
\end{figure}

\section{Cavity design} \label{sec:design}
\subsection{Structure and scheme}
Figure \ref{fig:design}(a) depicts the design of our PCN. The system is based on a Si slab with a refractive index of $n_{\rm Si} = 3.47$ and thickness $t$. The PhC here is a triangular lattice of circular air holes of radius $R_0$ and lattice constant $a$. Triangular-lattice PhC slabs are widely used in experiments because they have large photonic band gaps for TE-like modes. The lack of a single hole acts as a point defect and hence forms an H1 nanocavity, which is the simplest structure of PCNs that take over the ${\rm C_{6v}}$ symmetry of the PhC. The six holes closest to the defect, which are colored red in the figure, have a smaller radius $R_1$ than that of the background PhC ($R_1 < R_0$). This innermost layer of holes is also shifted radially away from the lattice points by a distance $s_1$. The second innermost hole layer comprises the twelve holes located one layer outward from the innermost ones and is drawn in orange. It is also translated in the radial direction so that it keeps the regular hexagonal alignment and its half diagonal is increased by a distance $s_2$. In addition, it's holes are of the same radius $R_0$ as those of the PhC.

We computed the complex eigenfrequencies $f$ of the hexapole eigenmode for various cases by using the finite element method on a commercial solver (COMSOL Multiphysics \cite{comsol}). With the defect center defined as the coordinate origin, the system had 11 and 14 layers of holes in the $\pm x$ and $\pm y$ directions, respectively. A rectangular air region with a height of 3 {\textmu}m was placed on each side of the slab. A scattering boundary condition for plane waves is applied to every border of the computational domain. The $x$-$y$ and $y$-$z$ planes were set as perfect magnetic and electrical conductors, respectively, for reducing the computational cost. Any changes to these simulation conditions are noted in what follows. The theoretical $Q$ factor is given by $Q_{\rm th} = {\rm Re} f / (2 {\rm Im} f)$.

Figure \ref{fig:design}(b) shows the $z$ component of the magnetic fields ($H_z$) of the hexapole mode along the $x$-$y$ plane. This mode is TE-like and thus characterized by $H_z$. It is also an eigenmode for the ${\rm C_6}$ rotation operator with an eigenvalue of $-1$. Such an odd parity of a symmetric two-dimensional multipole contributes to destructive interference in $H_z$ along the $z$ direction corresponding to $\Gamma$ point \cite{Johnson2001,Notomi2004}. This feature suppresses radiation loss based on the transverse electric field components $(E_x, E_y)$, as they are linked to $H_z$ through the Maxwell equations. Thus, structural modulation maintaining the lattice-matched rotational symmetry is essential to achieving an ultrahigh $Q$ factor of the hexapole mode. The other ${\rm C_6}$-symmetric eigenmode of this cavity is the monopole mode (not shown). It has an eigenvalue of $+1$ for the ${\rm C_6}$ operator and a far lower $Q_{\rm th} < 3000$ in our simulations.
\begin{figure}[h!]
	\centering\includegraphics[width=13cm]{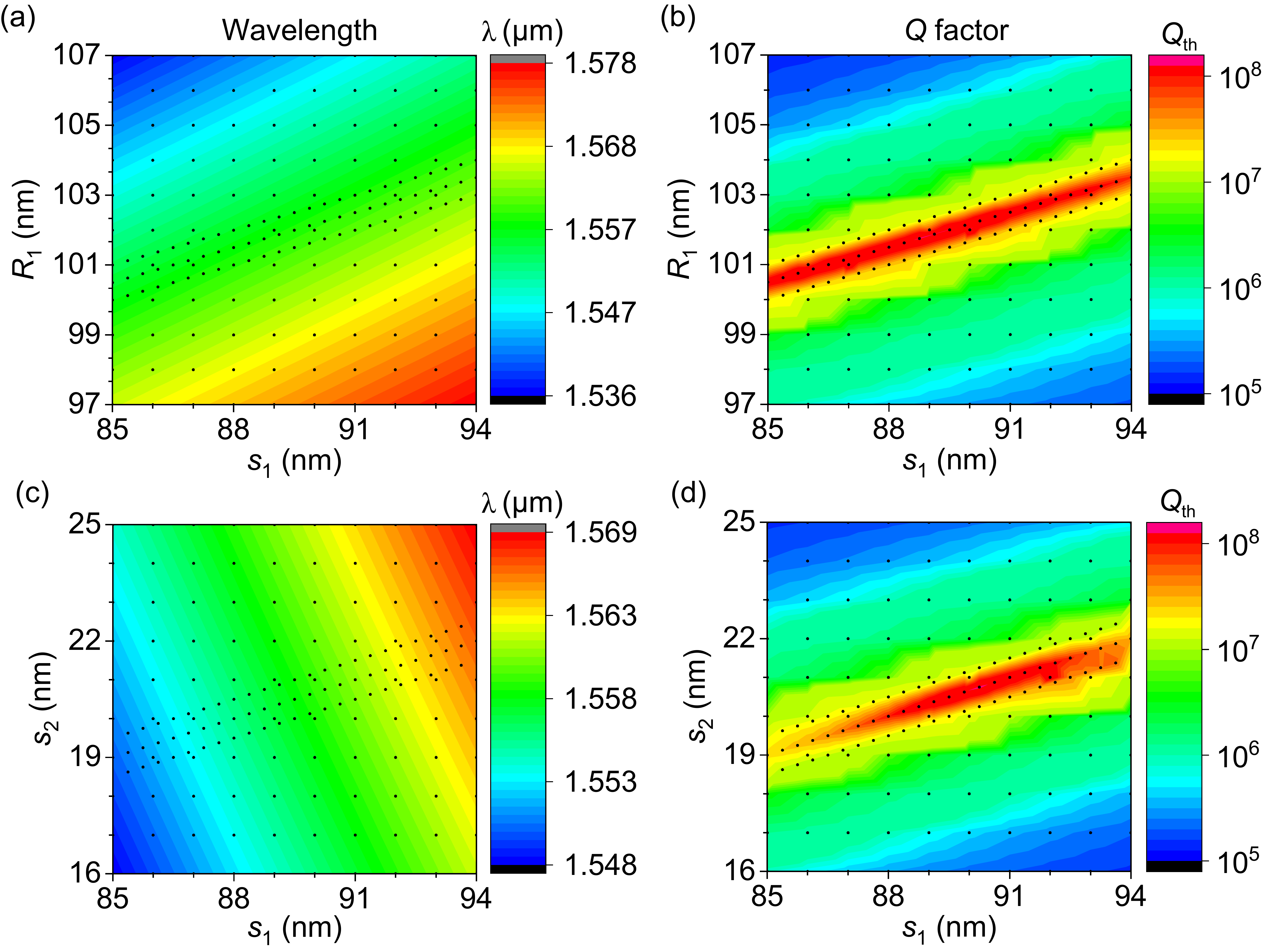}
	\caption{Dependence of (a) resonant wavelength ($\lambda$) and (b) theoretical $Q$ factor ($Q_{\rm th}$) of the hexapole mode on $s_1$ and $R_1$ for $s_2 = 20.5 \ {\rm nm}$. (c) $\lambda$ and (d) $Q_{\rm th}$ dependent on $s_1$ and $s_2$ for $R_1 = 102 \ {\rm nm}$. Black dots represent sample points in the simulation. The data among the points are linearly interpolated. A band of parameter conditions for $Q_{\rm th} > 10^8$ appears. $R_0 = 131 \ {\rm nm}$, $a = 426 \ {\rm nm}$, and $t = 250 \ {\rm nm}$.} \label{fig:designed wl and Q}
\end{figure}

As illustrated in Fig. \ref{fig:design}(a), our design uses only four parameters $(R_0, R_1, s_1, s_2)$ to improve the $Q$ factor, unlike recent designs based on costly optimizations of many variables \cite{Minkov2017,Asano2018,Shibata2021,Vasco2021}. $R_0$ determines the filling factor of the PhC, which is related to its photonic band gap and thus the in-plane modal confinement. $R_1$, $s_1$ and $s_2$ affect the local modal properties. The lattice constant $a$ can be varied to adjust the resonant wavelengths of the simulated modes to telecom ones around 1.55 {\textmu}m.

\subsection{Resonance properties versus hole shifts}
First, let us study the resonance characteristics of the mode for constant $R_0 = 131 \ {\rm nm}$, $a = 426 \ {\rm nm}$, and $t = 250 \ {\rm nm}$. Figure \ref{fig:designed wl and Q}(a) and (b) are two-dimensional color plots of the resonant wavelength $\lambda = c/{\rm Re}f$ and $Q_{\rm th}$ for isolated (unloaded) H1 PCNs depending on $s_1$ and $R_1$. Here, $c$ is the speed of light in vacuum and $s_2 = 20.5 \ {\rm nm}$. The plot of $\lambda$ indicates that a small $s_1$ and large $R_1$ squeeze the magnetic poles in Fig. \ref{fig:design}(b) and thus yield a short $\lambda$, whereas a large $s_1$ and small $R_1$ broaden the magnetic poles and thus increase $\lambda$. Remarkably, the $Q_{\rm th}$ plot exhibits a sequence of optimum points with $Q_{\rm th} > 10^8$ forming a linear band. Such a peak distribution indicates that there is an optimal polar width for every $\lambda$ that suppresses local scattering-induced radiation loss. There is a margin of about $\pm 1.5 \ {\rm nm}$ in $R_1$ and a wider one in $s_1$ from each optimum point to have a $Q_{\rm th} > 10^7$. The largest $Q$ factor here is $Q_{\rm th} = 1.43 \times 10^8$ for $(s_1, R_1) = (88.75 \ {\rm nm}, 101.75 \ {\rm nm})$. In units of $0.5 \ {\rm nm}$ for the parameters, $Q_{\rm th} = 1.41 \times 10^8$ for $(s_1, R_1) = (89.5 \ {\rm nm}, 102 \ {\rm nm})$ was obtained.

Figure \ref{fig:designed wl and Q}(c) and (d) depict the dependence of $\lambda$ and $Q_{\rm th}$ on $s_1$ and $s_2$ for $R_1 = 102 \ {\rm nm}$. There is a notable difference between Fig. \ref{fig:designed wl and Q}(a) and (c) in the directions of the iso-wavelength contours. This difference is due to negative correlation between the effect of $R_1$ and that of $s_2$; a larger $s_2$ results in a longer $\lambda$ because of the higher effective index of the cavity region. In contrast, Fig. \ref{fig:designed wl and Q}(b) and (d) appear to have more or less similar properties. As the mode wavelength increases with $s_1$, the optimal $s_2$ also becomes larger. $s_2$ can be used to dramatically improve $Q_{\rm th}$ because it introduces a gradual variation in the effective potential barrier of the PhC \cite{Song2005,Tanaka2008}. However, the trace of the extremely high $Q$ values in Fig. \ref{fig:designed wl and Q}(d) is nearly perpendicular to the contour lines in Fig. \ref{fig:designed wl and Q}(c), meaning that the conditions for a much improved $Q_{\rm th}$ are limited for each $\lambda$. The peak value of $Q_{\rm th}$ decreases for large and small $s_1$ because $R_1$ is fixed. Overall, a global optimization for $(R_1, s_1, s_2)$ enables us to find the continuous conditions for $Q_{\rm th} > 10^8$ in the parameter space. The best $Q_{\rm th}$ here is $1.46 \times 10^8$ for $(s_1, s_2) = (90.25 \ {\rm nm}, 20.75 \ {\rm nm})$.

\subsection{Modal properties}
\begin{figure}[h!]
	\centering\includegraphics[width=13cm]{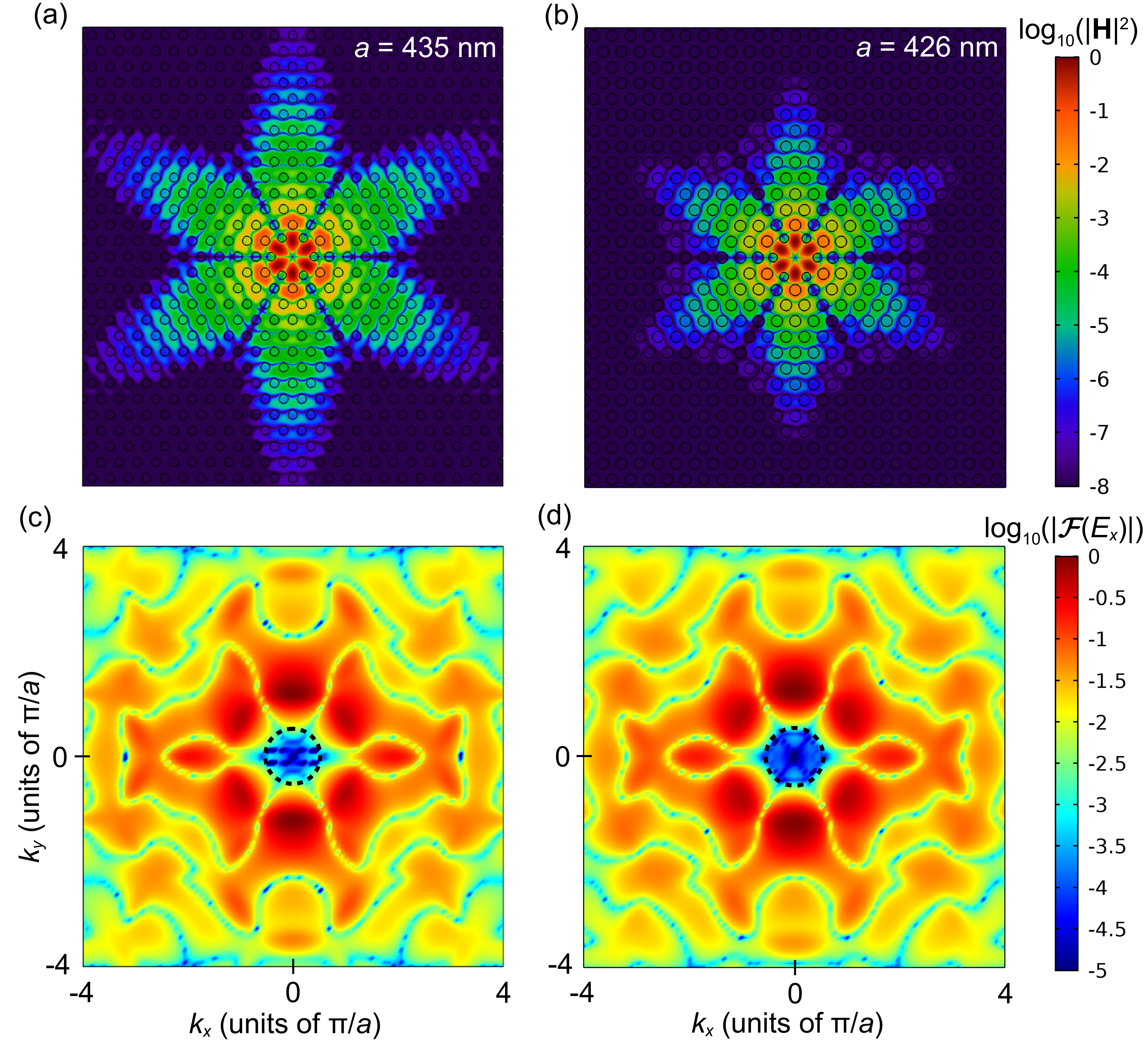}
	\caption{(a) Magnetic field intensity distribution in the logarithmic scale ($\log _{10}(|\mathbf{H}(\mathbf{r})|^2)$) for the hexpole nanocavity based on the previous work \cite{Kim2004} with $a = 435$ nm and $Q_{\rm th} = 2.0 \times 10^6$. (b) Same but for the hexpole mode designed in this study with $a = 435$ nm, $s_1 = 89.5$ nm, $R_1 = 102$ nm, and $Q_{\rm th} = 1.4 \times 10^8$, exhibiting more tightly confined in-plane evanescent fields than in (a). (c), (d) Absolute Fourier-space distributions of the $x$ components of the electric fields on a logarithmic scale ($\log _{10}(|\mathcal{F}(E_x(\mathbf{r}))|)$) for the eigenmodes corresponding to (a) and (b), respectively. (d) has significantly reduced radiative components inside the light line that is marked by the black dashed curve.} \label{fig:modal shapes}
\end{figure}

Next, let us compare the modal shapes in real and reciprocal spaces of the design with $Q_{\rm th} > 10^8$ and that in the previous study. Figure \ref{fig:modal shapes}(a) and (b) show the spatial magnetic intensity distributions on a common logarithmic scale ($\log _{10}(|\mathbf{H}(\mathbf{r})|^2)$) along $z = 0$ for hexapole modes with $Q_{\rm th} = 2.0 \times 10^6$ and $1.4 \times 10^8$, respectively. The PCN shown in (a) is based on Ref. \cite{Kim2004} and does not include $s_2$ in its design with $R_0 = 109 \ {\rm nm}$, $R_1 = 100 \ {\rm nm}$, $s_1 = 78 \ {\rm nm}$, $a = 435 \ {\rm nm}$, and $t = 220 \ {\rm nm}$. The other PCN in (b) corresponds to $(s_1, R_1) = (89.5 \ {\rm nm}, 102 \ {\rm nm})$ in Fig. \ref{fig:designed wl and Q}(a) and (b). A sizable portion of (a) has evanescent fields with relative intensities of about $10^{-4}$, and visible components with intensities over $10^{-8}$ reach the boundaries of the entire geometry. In comparison, the optimal mode shown in (b) obviously decays faster from the center. This means that the current design provides stronger in-plane light confinement.

Figure \ref{fig:modal shapes}(c) and (d) depict the Fourier transforms of the $x$ component of the electric fields on a logarithmic scale ($\log _{10}(|\mathcal{F}(E_x(\mathbf{r}))|)$) along $z = 0$ for the hexapole modes in Fig. \ref{fig:modal shapes}(a) and (b). Transverse electric field components lying within the LC measure the magnitude of radiation loss, because they can directly couple with radiative plane waves \cite{Srinivasan2002,Nakamura2016}. As shown in Fig. \ref{fig:modal shapes}(c), the previously designed mode has relative Fourier amplitudes of about $10^{-2.5}$ distributed in the LC defined by the black dashed circle. In stark contrast, the radiative field amplitudes are suppressed over the entire LC for the optimized mode shown in Fig. \ref{fig:modal shapes}(d). Their maximum value is about one order of magnitude smaller than that of Fig. \ref{fig:modal shapes}(c), confirming an improvement in the $Q$ factor due to the reduction of the radiation flux. A similar trend is seen in the case of $E_y$. These modal properties also support the discussion on Fig. \ref{fig:designed wl and Q}(b) and (d).

The standard Purcell mode volume $V$ for PCNs is given by \cite{Painter1999}
\begin{equation}
	V = \frac{\int \epsilon(\mathbf{r}) |\mathbf{E}(\mathbf{r})|^2 d^3 \mathbf{r}}{\max \{\epsilon(\mathbf{r}) |\mathbf{E}(\mathbf{r})|^2 \}}. \label{eq:mode volume}
\end{equation}
This definition is accurate in estimating the Purcell effect for high-$Q$ cavities and has been used for comparison purposes in the literature. Interestingly, the effective volume $V_{\rm opt} = 0.72 (\lambda/n)^3$ for the mode with $Q_{\rm th} = 1.4 \times 10^8$ is larger by $9\%$ than that of the previously studied one, $V_{\rm p} = 0.66 (\lambda/n)^3$. The electric energy densities of hexapole modes tend to concentrate mostly on the sides of the innermost air holes. However, the optimized mode distributes more electric energy around the point defect than the mode based on Ref. \cite{Kim2004} because of the potential modulation by $s_2$. Thus, it has a reduced maximum energy density or denominator in Eq. (\ref{eq:mode volume}).

This result shows that we can dramatically improve $Q_{\rm th}$ of the hexapole mode without sacrificing its small $V$. $V_{\rm opt}$ here is comparable with those of optimized L3 PCNs without hole radius modulation \cite{Nakamura2016,Shibata2021}, while the hexapole mode has a larger $Q_{\rm th}$. Thus, our H1 PCNs can be expected to have $Q_{\rm exp}$ values as high as those ones. In addition, our optimal $Q_{\rm th}/V_{\rm opt} = 1.9 \times 10^8 (n/\lambda)^3$ is slightly better than another L3 nanocavity with $Q_{\rm th}/V = 1.7 \times 10^8 (n/\lambda)^3$ ($Q_{\rm th} = 1.9 \times 10^8$ and $V = 1.1 (\lambda/n)^3$) designed by the particle-swarm algorithm \cite{Vasco2021}.

In summary, we showed designs of H1 PCNs based on a manual or brute-force search for extremely high-$Q$ hexapole modes. By focusing on the case for a constant $R_0$, we found a series of conditions for $Q_{\rm th} > 10^8$ with just three major optimization parameters $(R_1, s_1, s_2)$, thanks to the ${\rm C_6}$ symmetry of the mode. Introduction of an optical potential modulation with $s_2$ resulted in improved light confinement of the optimized mode in both the in-plane and out-of-plane directions. This point will be examined quantitatively in Sec. \ref{sec:optimization}.
\begin{figure}[h!]
	\centering\includegraphics[width=13.2cm]{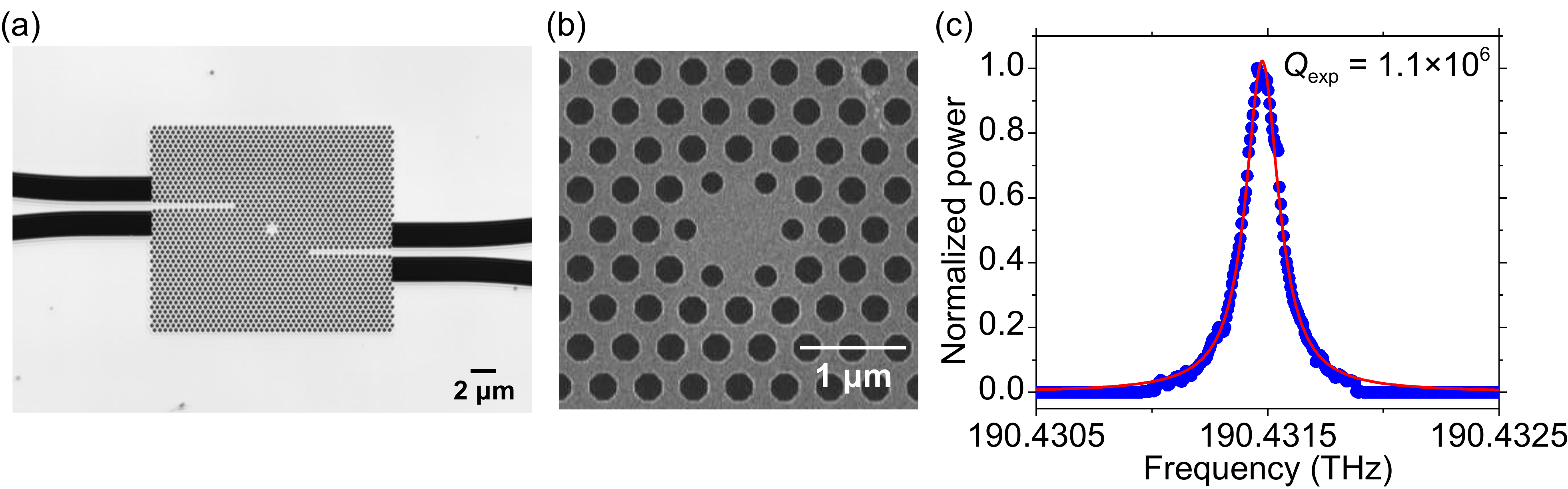}
	\caption{(a) Laser scope image of a sample with $d = 5 \sqrt{3} a$. The input and output Si waveguides are broadened and extended to both sides of the sample chip and coupled with lensed fibers. (b) Close-up SEM image of H1 PCN with $a = 434$ nm. Typical radii of the small and large air holes are estimated as $R_{\rm 1,s} \approx 106.8$ nm and $R_{\rm 0,s} \approx 133.1$ nm. (c) Transmission spectrum of sample with $a = 434$ nm and $s_1 = 99.5$ nm. The Lorentzian curve colored red matches the experimental data shown as blue points and indicates that the cavity has a loaded $Q_{\rm exp}$ of $1.1 \times 10^6$.} \label{fig:devices}
\end{figure} 

\section{Experimental result and numerical analysis}\label{sec:experiment}
\subsection{Sample fabrication and measurement}
We fabricated Si H1 PCNs of our design for an experimental demonstration. The sample structures were patterned by electron beam (EB) lithography on a positive EB resist coated on a silicon-on-insulator (SOI) wafer. The mask pattern was projected to the Si film with a nominal thickness of 250 nm by inductively coupled plasma etching. The buried oxide (BOX) layer beneath the PhCs was removed by wet etching with buffered hydrogen fluoride to obtain air-bridged samples. After the above device processes were completed, the wafer was cleaved so that the size of each sample chip was 5 mm $\times$ 15 mm. 

Figure \ref{fig:devices}(a) is a laser scope image of a PCN sample. The H1 cavity was butt-coupled (loaded) with two W1 PhC waveguides, each of which had a width of $W_0 = \sqrt{3}a$ based on the removal of a single row of air holes. The spatial interval $d$ between the cavity and them varied with the samples, and ones with $d = 5 \sqrt{3} a$ exhibited ultrahigh-$Q$ resonances. Each W1 waveguide was broadened by 100 nm at either end of the PhC by shifting five pairs of air holes on the sides outward with a stepwise increment of 20 nm. Consequently, they were efficiently coupled with air-suspended wire waveguides with a width of $W_0$. These optical channels were extended farther and connected to 8 {\textmu}m-wide slab waveguides that were supported by the BOX layer and led to the edges of the chip.

A close-up scanning electron microscope (SEM) image of an H1 nanocavity is shown as Fig. \ref{fig:devices}(b). Typical radii for the innermost and second innermost hole layers of the resist mask were estimated as $R_{\rm 1,m} \approx 102.8 \ {\rm nm}$ and $R_{\rm 0,m} \approx 130.4 \ {\rm nm}$, respectively, which were close to the condition for $Q_{\rm th} > 10^8$ found in Fig. \ref{fig:designed wl and Q}. However, the radii of the fabricated samples became somewhat bigger in the etching process: $R_{\rm 1,s} \approx 106.8 \ {\rm nm}$ and $R_{\rm 0,s} \approx 133.1 \ {\rm nm}$. We prepared PCN chips with five distinct lattice constants, $a = 418, 422, 426, 430, 434 \ {\rm nm}$. For the evaluations, we focused on the one with $a = 434 \ {\rm nm}$, because it best compensated for the discrepancies in hole radii between the design and fabrication.

We performed transmission measurements on each sample chip by placing it on a metallic stage whose temperature was maintained at 25${\rm ^\circ C}$ by a Peltier element and a PID controller. Tapered optical fibers were carefully aligned by using three-axis nano-positioners equipped with fiber holder stages, so that they were coupled with the slab waveguides at both ends of the chip and hence formed a measurement channel. The typical coupling loss per such interface was about 10 dB. A coherent transverse electric (TE) polarized light from a tunable laser was injected into each sample. The output was detected by a power meter synchronized with the wavelength sweep of the laser. The transverse magnetic (TM) field components of the input and output signals were filtered out by fiber polarizers. The entire system was based on polarization-maintaining fibers.

We prepared and measured a pair of H1 nanocavity samples with nominally the same structure for each of $s_1$; namely the shifts of the innermost holes varied from 89.5 to 101.5 nm in units of 1 nm. All these 26 samples had $s_2 = 20.5 \ {\rm nm}$ and $d = 5 \sqrt{3} a$. A transmission spectrum of an H1 nanocavity with $s_1 = 99.5 \ {\rm nm}$ is plotted in Fig. \ref{fig:devices}(c). The experimental data shown as blue points match the Lorentzian curve (colored red) obtained by a least squares fitting. The peak frequency (wavelength) was 190.4315 THz (1575.370 nm), and the linewidth of the best-fit curve was 173.8 MHz. These values give an experimental loaded $Q$ factor of $Q_{\rm exp} = 1.1 \times 10^6$. Here, we have excluded any arbitrariness in determining $Q_{\rm exp}$ of the measured resonance with discrete data points. The input power was attenuated so that thermal linewidth broadening and nonlinearity would be avoided. In this case, however, the detection power around resonance tails tended to be slightly reduced, as indicated by its visible drop near 190.4319 THz. This is because the power meter had a limited dynamic range with a minimum detectable power of -80 dBm.

We can certainly identify this resonance to be the hexapole mode, because the other cavity modes typically have $Q_{\rm th} < 20000$ in our simulations and their wavelength spacing with respect to the ultrahigh-$Q$ peak is 30 nm or larger.

\subsection{Measured wavelengths and quality factors of H1 PCNs}
\begin{figure}[h!]
	\centering\includegraphics[width=13cm]{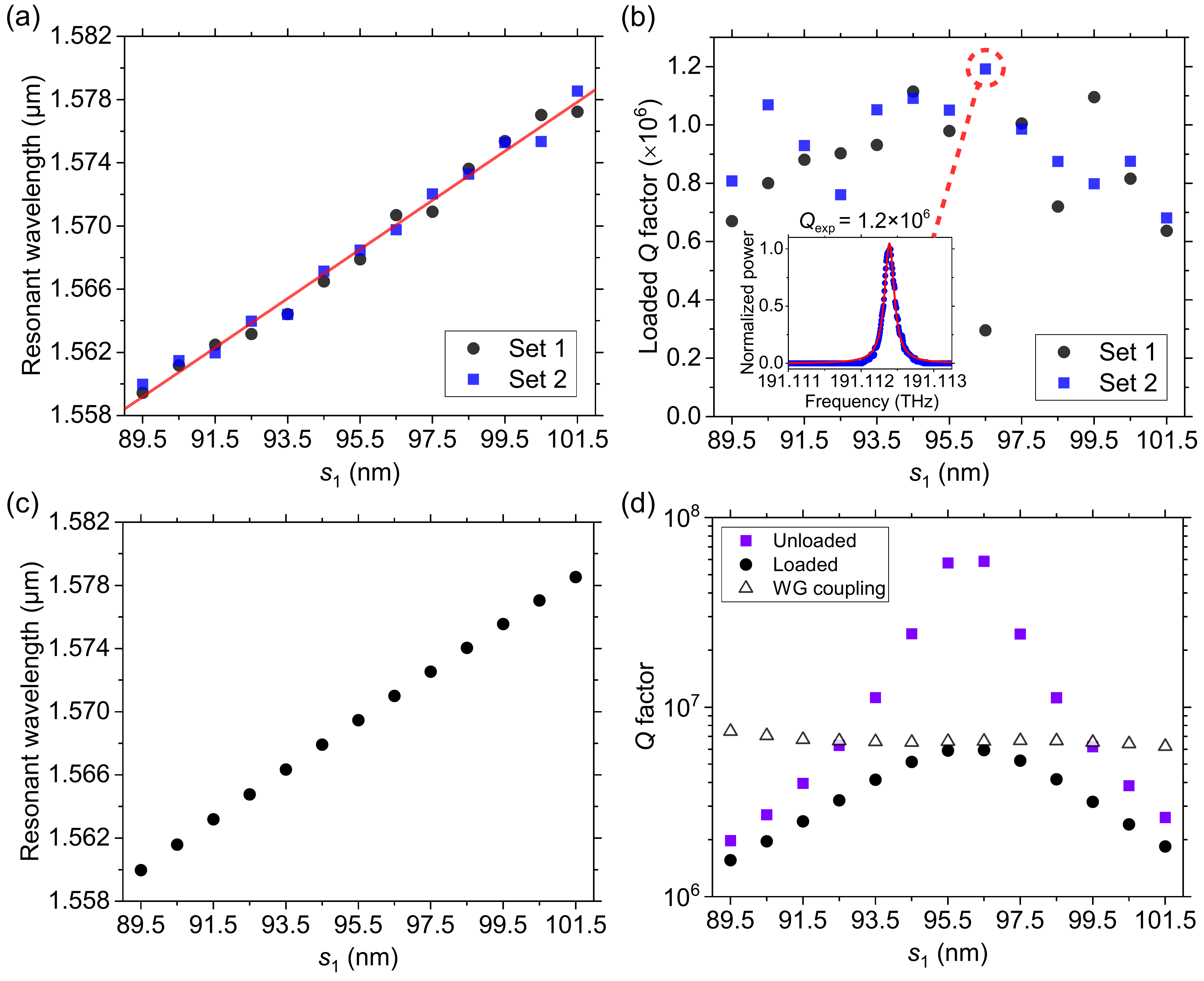}
	\caption{(a) Dependence of measured $\lambda$ on $s_1$ for two nominally duplicate sets of H1 PCN samples with $a = 434$ nm, $s_2 = 20.5 \ {\rm nm}$, and $d = 5 \sqrt{3} a$. The grouping of the samples into sets is based on their positions relative to the front edge of chip (the samples in set 1 are closer to the edge). The red line is a linear regression of the experimental data. (b) Loaded $Q$ factor ($Q_{\rm exp}$) as a function of $s_1$ for the two sample sets. The inset is the transmission spectrum for the best sample that had $Q_{\rm exp} = 1.2 \times 10^6$ and $s_1 = 96.5$ nm. (c) Simulated $\lambda (s_1)$ for $a = 434$ nm, $t = 241$ nm, $R_1 = 106$ nm, $R_0 = 134$ nm, and $s_2 = 20.5$ nm, which agrees well with the experimental data. (d) Simulated $Q$ factors for the same parameters on a semi-logarithmic scale. Squares show results for unloaded samples ($Q_{\rm th}$), while dots are for loaded ones ($Q_{\rm th,L}$) including two W1 PhC waveguides with $d = 5 \sqrt{3} a$ that radiate out the light. Triangles show the $Q$ factors $Q_{\rm WG}$ due to the losses by the waveguides.} \label{fig:measurement_result}
\end{figure}

Figure \ref{fig:measurement_result}(a) presents the dependence of the measured resonance wavelengths $\lambda$ of the hexapole modes on $s_1$. To show the correspondence between the data of $\lambda$ and $Q_{\rm exp}$, we divided the samples into two sets according to their positions, so that each sample in set 1 is closer to the front edge of the chip than its counterpart in set 2 with the same $s_1$. It can be clearly seen that $\lambda$ is positively correlated with $s_1$, as predicted in Fig. \ref{fig:designed wl and Q}(a) and (c). The variation in $\lambda$ within pairwise samples for each $s_1$ is so weak that a linear regression of the entire data, shown by the red line, reproduces their average trend. The slope of the regression line is $1.55 \pm 0.032$ nm ($\lambda$) / nm ($s_1$), and its coefficient of determination is $R^2 = 0.990$. 

Here we define the difference in resonant wavelength between set 1 and 2 as $\Delta \lambda (s_1) = \lambda_{1}(s_1) - \lambda_{2}(s_1)$, where $\lambda_{1}(s_1)$ and $\lambda_{2}(s_1)$ are the wavelengths of the samples with $s_1$ in set 1 and 2, respectively. $\Delta \lambda$ for all $s_1$ in Fig. \ref{fig:measurement_result}(a) are calculated, and then their standard deviation is found to be $\sigma_{\Delta \lambda} = 0.848 \ {\rm nm}$. Because $\lambda_{1}(s_1)$ and $\lambda_{2}(s_1)$ ideally have the same value and their variations should stem from numerous independent and random processes during fabrication, we assume that they have no covariance. Thus, we can estimate the deviation in $\lambda$ to be $\sigma_{\lambda} = [ \sigma_{\Delta \lambda}^2/2 ]^{1/2} = 0.600 \ {\rm nm}$.

This result implies that our nanocavities have highly accurate hole positions. Although the obtained value of $\sigma_{\lambda}$ corresponds to a change solely in $s_1$ of 0.39 nm, in reality, there are other major factors that affect $\sigma_{\lambda}$, such as the hole radii, local Si slab thicknesses and surface roughness. In addition, the positioning accuracy of the electron beam used in patterning the resist mask is as small as 0.05 nm. Thus, undesired variations in hole positions, including those in $s_1$ and $s_2$, will be less significant in the actual samples.

The measured loaded $Q$ factors for the two sample sets are plotted in Fig. \ref{fig:measurement_result}(b) as a function of $s_1$. They exhibit a gentle peak centered around $s_1 = 94.5$ or 95.5 nm; $Q_{\rm exp}$ for these values of $s_1$ is significantly larger than that for $s_1 = 89.5$ and 101.5 nm. The best sample here belongs to set 2 and has $s_1 = 96.5$ nm and $Q_{\rm exp} = 1.2 \times 10^6$ with an estimated linewidth of 160.4 MHz. Its transmission spectrum is shown in the inset of Fig. \ref{fig:measurement_result}(b). Although the shape of the resonance is slightly asymmetric, it is still fitted by a Lorentzian function.

Eight samples out of 26 had $Q_{\rm exp} > 10^6$. Remarkably, they included ones with $s_1 = 90.5$ and 99.5 nm, namely off from the peak center. This trend implies that the $Q$ factors for these PCNs are much larger in theory but were reduced because of fabrication imperfections. The effect of disorder is also reflected in the outlier sample with a low $Q_{\rm exp} = 3.0 \times 10^5$ and $s_1 = 96.5$ nm in set 1.

\subsection{Simulation of measured samples}
We performed simulations by varying the structural parameters around those estimated from the SEM image. Figure \ref{fig:measurement_result}(c) shows the theoretical $\lambda$ as a function of $s_1$ for $a = 434$ nm, $t = 241$ nm, $R_1 = 106$ nm, $R_0 = 134$ nm, and $s_2 = 20.5$ nm. The theoretical values agree well with the experimental data. Although the simulation result is slightly convex upward, its average slope (1.55 nm ($\lambda$) / nm ($s_1$)) coincides with that of the experimental result. We emphasize that $R_1$ and $R_0$ here are consistent with the measured $R_{\rm 1,s}$ and $R_{\rm 0,s}$ within an error of a few nanometers, as expected for the current measurement. The value of $t$ is smaller than the nominal thickness 250 nm of the Si film, indicating that the PhC slabs were thinned down by the etching processes and/or that $n_{\rm Si}$ in the simulation is slightly smaller than that of the actual material.

Moreover, as shown in Fig. \ref{fig:measurement_result}(d), the corresponding theoretical $Q$ factors follow the trend seen in the experiment. The figure compares $Q_{\rm th}$ for the H1 PCNs with and without two W1 PhC waveguides with $d = 5 \sqrt{3} a$ extending to the right and left sides of the simulation domain where the fields are scattered out. The plots are on a semi-logarithmic scale, with the horizontal axis depicting steps of 1 nm. The loaded $Q$ factors, $Q_{\rm th,L}$, are the black dots, and the unloaded ones, $Q_{\rm th}$, are the purple squares. Both plots peak at $s_1 = 96.5$, where $Q_{\rm th,L} = 5.9 \times 10^6$ and $Q_{\rm th} = 5.9 \times 10^7$. The loaded hexapole mode for this condition has a theoretical modal volume of $V = 0.74 (\lambda/n)^3$. Thus, our best experimental sample is expected to have had $Q_{\rm exp}/V = 1.6 \times 10^6 (n/\lambda)^3$. 

The difference between $Q_{\rm th,L}$ and $Q_{\rm th}$ comes from the coupling with the environment via the waveguides. The impact of this coupling, $Q_{\rm WG}$, can be derived from the relation $1/Q_{\rm th,L} = 1/Q_{\rm th} + 1/Q_{\rm WG}$. The resultant values are plotted as the triangles in Fig. \ref{fig:measurement_result}(d). They exhibit a moderate variation with $s_1$ probably due to the group velocity dispersion of the waveguides and are about $Q_{\rm WG} = 6.6 \times 10^6$ around the peak of $Q_{\rm th}$. As a result, the intrinsic (unloaded) $Q$ factor of the optimum sample is estimated to be $Q_{\rm i} = [1/Q_{\rm exp} - 1/Q_{\rm WG}]^{-1} = 1.5 \times 10^6$. The correspondent $Q/V$ amounts to $Q_{\rm i}/V = 2.0 \times 10^6 (n/\lambda)^3$, which is comparable with those of PCNs without having their surface Si passivated with hydrogen \cite{Tanabe2007,Simbula2017,Lai2014,Dharanipathy2014}.

\subsection{Impact of varying hole radii}
We can see that $Q_{\rm exp}$ is still lower than $Q_{\rm th,L}$ and hence it is expected to be affected by reductive factors other than $Q_{\rm WG}$. A simple but realistic cause of extra loss is radiative scattering induced by random variations in the radii and positions of the air holes \cite{Hagino2009,Taguchi2011}. The hole radii can change on the atomic scale order because of stochastic processes in fabrication, such as in the EB exposure, resist development, and dry and wet etching. On the other hand, the EB shots are precisely aligned in our lithography process. Thus, the positions of the hole centers are mainly affected by the small and probabilistic anisotropy of etching or distortion in the shapes of the holes, part of which is also considered to impact the radii.

Here, we simulated samples with air holes just of varying radii to statistically evaluate the effect of fabrication imperfections on the $Q$ factor. The result estimates a dominant portion of the disorder-induced scatting loss denoted as $1/Q_{\rm scat}$. We used the parameters that reproduce $\lambda$ of the measured samples and set $s_1 = 96.5$ nm for $Q_{\rm th} = 5.9 \times 10^7$ without structural imperfections or PhC waveguides. The PEC boundary condition of the $y$-$z$ plane was removed so that the simulation explicitly included all the holes. The small and large holes were assumed to have random radii sampled from Gaussian distributions with means $R_1$ and $R_0$, respectively, and a common standard deviation (SD) of $\sigma_r$. The $Q$ factor obtained in each run is denoted as $Q_{\rm th,F}$ and satisfies $1/Q_{\rm th,F} = 1/Q_{\rm th} + 1/Q_{\rm scat}$. 
\begin{figure}[h!]
	\centering\includegraphics[width=13cm]{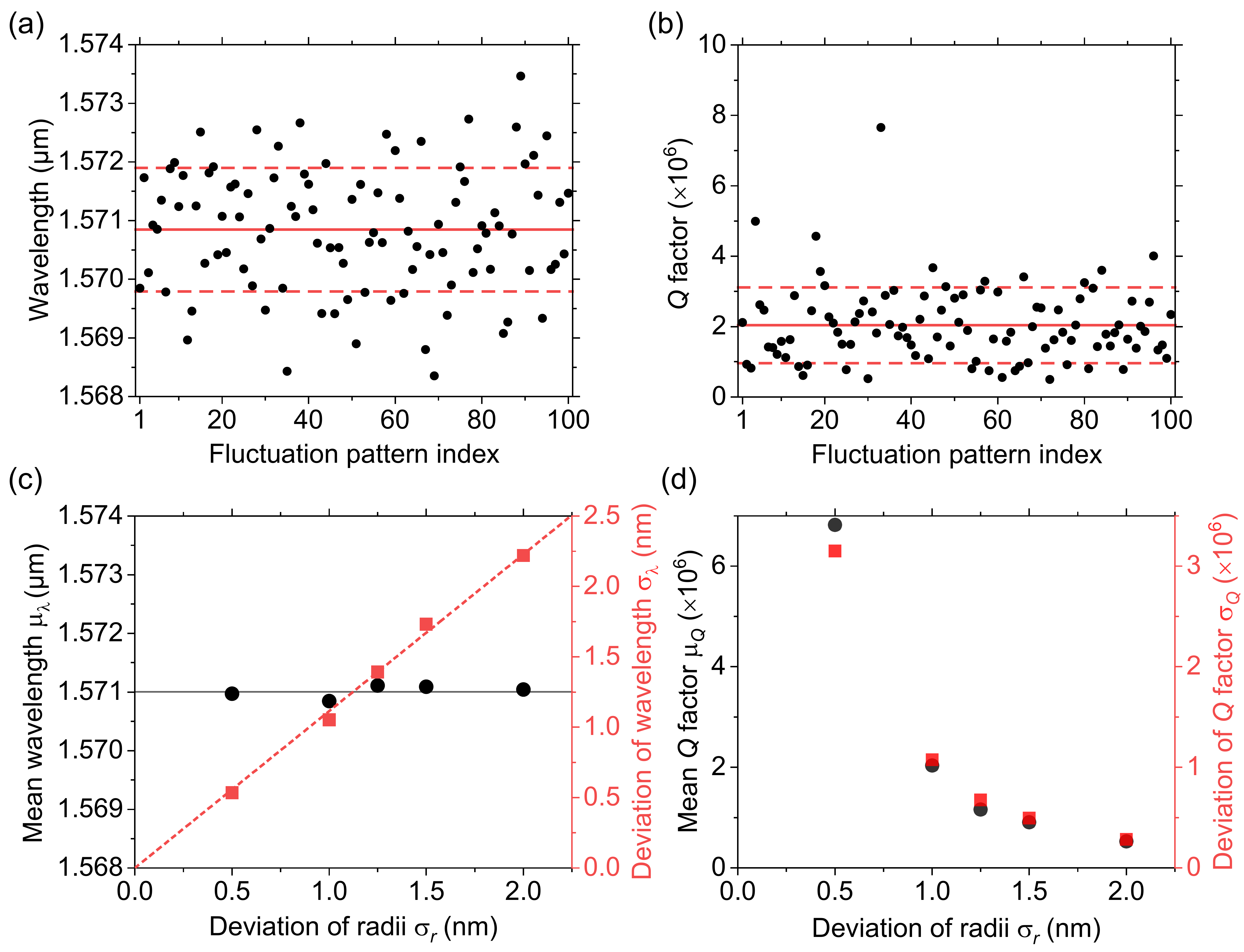}
	\caption{(a) Simulated resonant wavelengths and (b) unloaded $Q$ factors of H1 PCNs with 100 different random patterns of hole radii for $\sigma_r = 1.0$ nm. (c) Mean and standard deviation of the resonant wavelength $(\mu_{\lambda}, \sigma_{\lambda})$ and (d) those of the $Q$ factor $(\mu_{Q}, \sigma_{Q})$ of the random simulation for different $\sigma_r$. $\mu_{\lambda}(\sigma_r )$ converges at the result without any disorder shown as the black line, while $\sigma_{\lambda}(\sigma_r )$ grows linearly, as indicated by the regression line in red. Both $\mu_{Q}$ and $\sigma_{Q}$ are inversely proportional to $\sigma_r^2$. The approximate statistical properties of the scattering loss are given by Eqs. (\ref{eq:avginvQ}) and (\ref{eq:SDinvQ}). The mean $R_0$ and $R_1$ are 134 nm and 106 nm, respectively. The other parameters are the same as those used for Fig. \ref{fig:measurement_result}.} \label{fig:simulation_result_random}
\end{figure}

Figure \ref{fig:simulation_result_random}(a) and (b) show $\lambda$ and $Q_{\rm th,F}$ for 100 random patterns with $\sigma_r = 1.0$ nm. The data points of both plots look randomly scattered. The mean and SD of the resonant wavelengths are $(\mu_{\lambda}, \sigma_{\lambda}) =$ (1.57084 {\textmu}m, 1.052 nm) and those of the $Q$ factors are $(\mu_{Q}, \sigma_{Q}) = (2.3 \times 10^6, 1.07 \times 10^6)$. The wavelengths tend to be distributed symmetrically around $\mu_{\lambda}$, while the $Q$ factors are specifically high for some sample points, indicating distinct statistical properties.

We repeated the random simulations for different $\sigma_r$. The dependence of $(\mu_{\lambda}, \sigma_{\lambda})$ on $\sigma_r$ and that of $(\mu_{Q}, \sigma_{Q})$ are plotted in Fig. \ref{fig:simulation_result_random}(c) and (d), respectively. The mean wavelength for each $\sigma_r$ is mostly convergent at $\lambda = 1.5710$ {\textmu}m, which is obtained for the case of no disorder. The deviation in $\lambda$ grows proportionally with $\sigma_r$. The variance of the radii $\sigma_r ^2$ is directly related to that of the effective dielectric constant of the PhC slab via the filling fraction of the air holes. Thus, $\sigma_r$ affects the deviation of the effective index and has an approximately linear dependence on $\sigma_{\lambda}$. Its slope is estimated as $\sigma_{\lambda} / \sigma_r = 1.11$.

In contrast, both $\mu_{Q}$ and $\sigma_{Q}$ tend to be inversely proportional to $\sigma_r ^2$. As discussed in Ref. \cite{Hughes2005}, local variations in the dielectric constant affect the extra scattering rate and hence the loss. By subtracting $1/Q_{\rm th}$ from $1/Q_{\rm th,F}$ of the data, the approximate mean and SD of $1/Q_{\rm scat}$ are given by
\begin{eqnarray}
	\mu[1/Q_{\rm scat}] = 6.3 \times 10^{-7} \sigma_r ^2, \label{eq:avginvQ}\\
	\sigma[1/Q_{\rm scat}] = 3.3 \times 10^{-7} \sigma_r ^2,\label{eq:SDinvQ}
\end{eqnarray}
where $\sigma_r$ is measured in nanometers. Similar properties have been reported in multi-heterostructure nanocavities with variations in the positions and radii of the air holes \cite{Hagino2009,Taguchi2011}.

As mentioned in the discussion of Fig. \ref{fig:measurement_result}(a), the experimental data suggest $\sigma_{\lambda} = 0.600$ nm. This value corresponds to $\sigma_r = 0.54$ nm via the proportional relation between $\sigma_{\lambda}$ and $\sigma_r$. By substituting the value of $\sigma_r$ into Eqs. (\ref{eq:avginvQ}) and (\ref{eq:SDinvQ}), we obtain $\mu[1/Q_{\rm scat}] = 1.8 \times 10^{-7}$ and $\sigma[1/Q_{\rm scat}] = 9.6 \times 10^{-8}$, as the estimated statistical properties of the scattering loss for the measured samples. The resultant mean $Q_{\rm scat}$ is $5.4 \times 10^{6}$. We should emphasize that we did not underestimate $Q_{\rm scat}$ by neglecting inaccuracies in the hole positions. The variation in wavelength in the experiment is attributed solely to $\sigma_r$, and its entire impact is hence taken into consideration in obtaining $Q_{\rm scat}$.
\begin{figure}[h!]
	\centering\includegraphics[width=13cm]{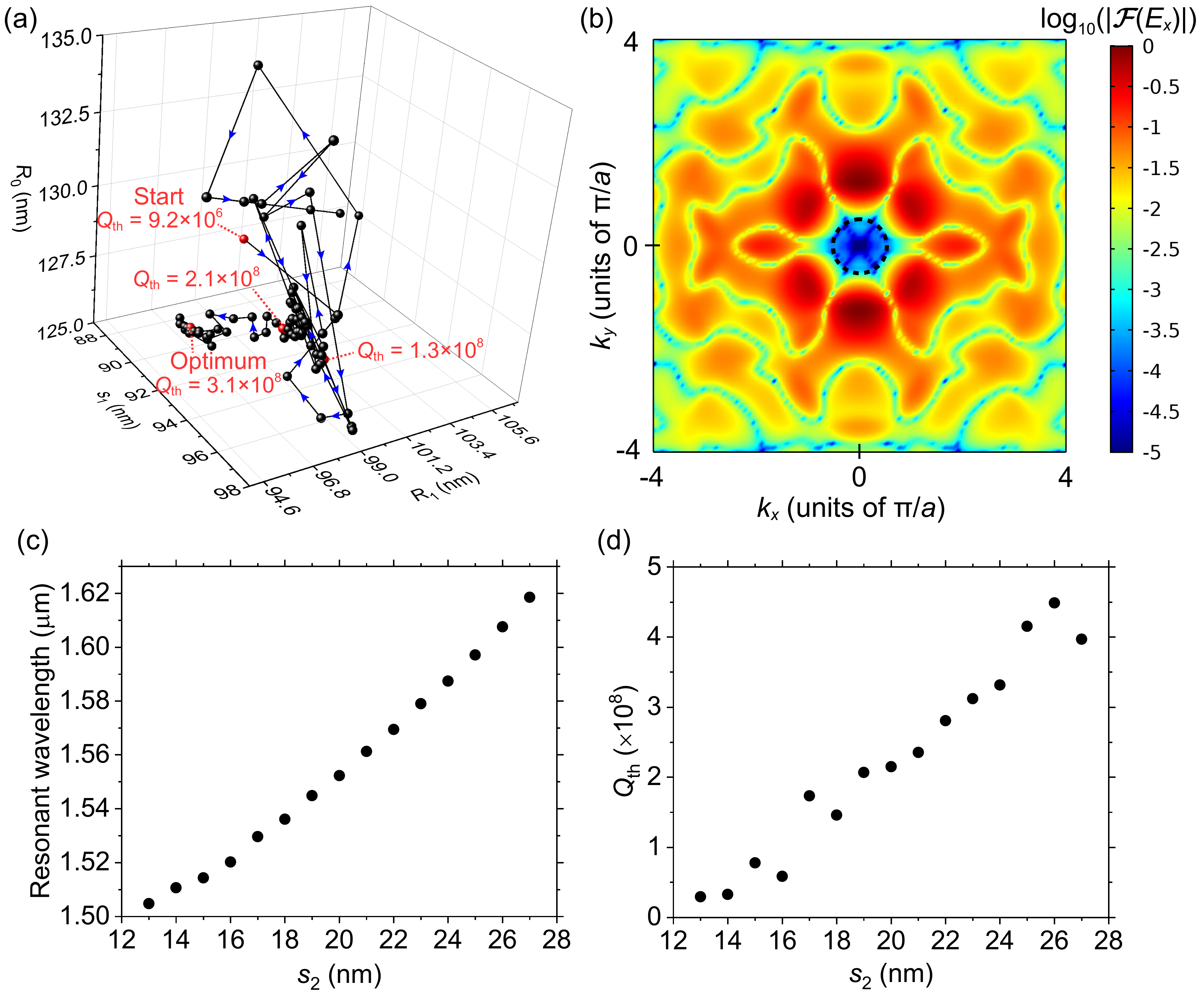}
	\caption{(a) Evolution of $(R_0, R_1, s_1)$ in the Nelder-Mead optimization of $Q_{\rm th}$ for $s_2 = 23$ nm. Blue arrows indicate the direction of the parameter variation. (b) $\log _{10}(|\mathcal{F}(E_x(\mathbf{r}))|)$ for the optimized hexapole mode for $s_2 = 23$ nm. The radiative component lying inside the LC is reduced, compared with Fig. \ref{fig:modal shapes}. The black dashed circle denotes the light line. (c) $\lambda$ and (d) $Q_{\rm th}$ of the optimized H1 PCNs for different $s_2$. Both of them tend to be positively correlated with $s_2$. We obtained $Q_{\rm th} = 4.5 \times 10^8$ for the optimized variables $(R_0, R_1, s_1) \approx (115.92 \ {\rm nm}, 90.258 \ {\rm nm}, 85.773 \ {\rm nm})$ for $s_2 = 26$ nm. Other fixed parameters are $a = 426$ nm and $t = 250$ nm.}
	\label{fig:optimization}
\end{figure} 

Because the mean $Q_{\rm exp}$ is $\mu[Q_{\rm exp}] \approx 10^{6}$ around the optimal condition, this result indicates the existence of further loss in the experiment with an average $Q$ factor of $(\mu[1/Q_{\rm exp}] - \mu[1/Q_{\rm scat}] - 1/Q_{\rm WG})^{-1} \approx 1.5 \times 10^{6}$. We attribute part of this loss to a slight amount of EB resist remaining on the sample. Considering that the laser scope comes into focus twice in scanning the surface, it is expected to form a very thin layer over the chip. This results in structural asymmetry in the out-of-plane direction and hence induces extra radiation loss, as is the case with samples fabricated on sacrificial layers. Its unevenness, which can be seen at the top right of Fig. \ref{fig:devices}(b) for example, could also be a source of scattering. We did not try to remove the resist layer from the chip, because such a process unavoidably thins down the Si layer and thus alters the dependence of the resonance properties on $s_1$. The sample quality will be improved in future studies.

\section{Automated optimization}\label{sec:optimization}
Recent studies have used various automated optimization algorithms to achieve high theoretical $Q$ factors in PCNs \cite{Minkov2014,Lai2014,Minkov2017,Asano2018,Shibata2021,Vasco2021}. We used the built-in optimization module of COMSOL Multiphysics and found that the performance of the H1 PCN can further be improved. Here, we chose the Nelder-Mead method \cite{Nelder1965}, which prepares a symplex in a parameter space and repeats its update based on the reflection, expansion, contraction, or shrink process, depending on the value of the function $F$ to be optimized. This scheme does not use any gradient or assume any approximate form of the function. Thus, it is expected to work regardless of the actual landscape of $F$. We fixed $s_2$ and obtain a maximal $Q_{\rm th}$ by varying $R_0$, $R_1$ and $s_1$ in each optimization run, namely $F = Q_{\rm th} (R_0, R_1, s_1)$.

Figure \ref{fig:optimization}(a) shows the evolution of the parameters in the optimization for $s_2 = 23$ nm, $a = 426$ nm, and $t = 250$ nm. Here, the initial point was set as $(R_0, R_1, s_1) = (128.3 \ {\rm nm}, 99.5 \ {\rm nm}, 89.4 \ {\rm nm})$ with $Q_{\rm th} = 9.2 \times 10^6$. The variables undergo substantial changes at steps in the early stage of the operation. The state passes through a condition for $Q_{\rm th} > 10^8$ and is then bound in a region of suboptimal points with $Q_{\rm th} < 2 \times 10^8$. After a while, however, the algorithm finds a direction in which $Q_{\rm th}$ is improved beyond $2 \times 10^8$. It eventually settles at $(R_0, R_1, s_1) \approx (125.18 \ {\rm nm}, 97.421 \ {\rm nm}, 89.024 \ {\rm nm})$ exhibiting the optimum objective, $Q_{\rm th} = 3.1 \times 10^8$. The normalized absolute Fourier amplitudes of $E_x$ for this optimal mode are depicted on a logarithmic scale in Fig. \ref{fig:optimization}(b). Compared with Fig. \ref{fig:modal shapes}(d), the domain with the relative amplitudes below $10^{-5}$ in the LC is doubly extended in the $k_x$ direction. This feature confirms that the light confinement of this H1 PCN is stronger than that of the manually designed ones shown in Sec. \ref{sec:design}.

We repeated the optimization routine with different values of $s_2$, which is the additional factor not in the former design examined in Fig. \ref{fig:modal shapes}(a) and (c). To understand quantitatively the impact of $s_2$, we plot the dependences of $\lambda$ and $Q_{\rm th}$ of the optimized PCN in Fig. \ref{fig:optimization}(c) and (d). The resonant wavelength is monotonically red-shifted as $s_2$ increases. Accordingly, a larger $s_2$ results in a higher optimal $Q$ factor. We find that $Q_{\rm th} = 4.5 \times 10^8$ for $s_2 = 26$ nm, which is more than a hundred-times the values in the previous reports \cite{Kim2004,Minkov2014}. Remarkably, the optimized mode also has a small volume of $V_{\rm opt} = 0.71 (\lambda/n)^3$, and thus its $Q/V$ is as large as $Q_{\rm th}/V_{\rm opt} = 6.3 \times 10^8 (n/\lambda)^3$. This result confirms the striking contribution of the gradual variation in the optical potential introduced by $s_2$ to $Q_{\rm th}$, as mentioned in Sec. \ref{sec:design}.

The optimal structural parameters vary greatly with $s_2$. We obtained $(R_0, R_1, s_1)\approx (144.23 \ {\rm nm}, 111.61 \ {\rm nm}, 86.020 \ {\rm nm})$ and $(115.92 \ {\rm nm}, 90.258 \ {\rm nm}, 85.773 \ {\rm nm})$ for $s_2 = 13$ nm and 26 nm, respectively. $R_0$ and $R_1$ tend to be negatively correlated with $s_2$ and $\lambda$, while $s_1$ oscillates gently between 82 nm and 92 nm with respect to $s_2$. Optimization with more parameters such as $(R_0, R_1, s_1, s_2, a)$ might result in a better $Q_{\rm th}$. In that case, however, the parameter space would become larger and contain more local minima of $Q_{\rm th}$. Thus, the computation would be much harder in terms of both its convergence and the probability of finding a good solution. We leave that consideration out of the scope of this study.

\section{Discussion}\label{sec:discussion}
Experimental $Q$ factors of PCNs are generally limited by many kinds of defects. Discussing their impact will allow us to predict how high $Q_{\rm exp}$ could be made in a real PCN device.

A major cause of the reduction of $Q$ factors is structural imperfections. In our result, the variations in $\lambda$ and $1/Q_{\rm th,F}$ were attributed to those in the hole radii, and $\sigma_r = 0.54$ nm and $\mu[1/Q_{\rm scat}] = 1.8 \times 10^{-7}$ were obtained. A groundbreaking report by Asano et al. on multi-heterostructure PCNs \cite{Asano2017}, including one with $Q_{\rm exp} = 1.1 \times 10^7$, considered the same deviation $\sigma_{\rm hole}$ in both the positions and radii of the air holes. They estimated $\sigma_{\rm hole}$ to be 0.25 nm and the corresponding $\mu[1/Q_{\rm scat}]$ to be $4.7 \times 10^{-8}$ for their PCN samples. A monolayer of Si is about 0.135-nm-thick and an air hole has two side walls in the radial direction. Thus, $\sigma_{\rm hole} = 0.25$ nm seems to indicate that the etching process just leaves the uncertainty at the level where a single atomic layer is removed or not at every Si surface, including the resultant hole displacement. Both Eq. (\ref{eq:avginvQ}) and the dependence of $\mu[1/Q_{\rm scat}]$ on $\sigma_{\rm hole}$ in Ref. \cite{Asano2017} are quadratic equations and have similar coefficients. Even though $\sigma_r$ and $\sigma_{\rm hole}$ of the two PCNs can be reduced to the monolayer level ($= 0.135$ nm), a dimensionless loss of about $\mu[1/Q_{\rm scat}] \approx 10^{-8}$ remains. This implies that it is hard to achieve $\mu[Q_{\rm scat}] > 10^{8}$ for PCNs.
 
Another limiting factor is the formation of surface oxidation layers on Si. Every Si/SiO$_x$ interface has a few kinds of surface states whose spectral densities of states are within the band gap of Si \cite{Yamashita1996}. They exhibit optical absorption at telecommunication wavelengths ($\approx 0.8$ eV) and are known to significantly increase loss in Si photonic devices \cite{Borselli2006}. This detrimental effect can be circumvented by passivating Si surfaces with hydrogen via HF etching \cite{Yablonovitch1986,Takahagi1988}. However, the Si-H termination is not stable and the surfaces hence suffer from natural oxidation in ambient conditions. Thus, a combination of this process and subsequent measurement of the samples in an inert gas-purged chamber seems to be needed in order to achieve $Q_{\rm exp} > 10^7$ \cite{Asano2017}. For heterostructure PCNs with oxide layers \cite{Sekoguchi2014}, the inverse of the $Q$ factor based on absorption ($1/Q_{\rm abs}$) was estimated to be about $1/(7 \times 10^6) = 1.43 \times 10^{-7}$, and a large part of it seemed to stem from the surface states. Although water molecules that adhere to sample surfaces also induce absorption loss, their impact appears to be an order of magnitude smaller. Repeating the formation and removal of SiO$_x$ layers can also reduce the surface roughness and hence suppress extra scattering loss \cite{Lee2001,Sparacin2005}. Performing such a process on the bottom surface of Si may also be helpful in removing dopant contamination that could concentrate around the interface between the Si and BOX layers \cite{Ling1992,Asano2017}.

Overall, the $Q_{\rm exp}$ achievable for practical PCNs in air seems to be limited to below $10^7$; with the hydrogen passivation $Q_{\rm exp}$ may reach on the order of $10^7$. Because PCNs can have such a high $Q/V$ coefficient, we should mention that they would also be subject to fluctuations in the refractive index caused by thermal noise, which induce their linewidth broadening \cite{Panuski2020}. Although PCNs are not so affected by ambient temperature, thermal noise may become a problem when they absorb the injected light. Our experiment showed a symptom of the linewidth broadening, when the measured transmission power exceeded 1 nW. This feature is attributed to heat, since it appears as a precursor of bistable transmission based on thermo-optic nonlinearity. A similar result was seen in a previous report \cite{Tanabe2007_2}. PCNs with larger $Q_{\rm exp}$ than ours might need a smaller probe power to avoid it. In that case, a time-resolved ("ring-down") measurement with a pulsed excitation might be useful \cite{Tanabe2007_3}.

\section{Conclusion}\label{sec:conclusion}
The theoretical and experimental $Q$ factors of our hexapole H1 PCNs were $Q_{\rm th} > 10^8$ and $Q_{\rm exp} > 10^6$. Thanks to the $C_{\rm 6}$ symmetry of the hexapole mode, our design required optimization of only four structural modulation parameters. Bands of valid conditions for $Q_{\rm th} \gtrapprox 10^8$ were found in both the $(s_1, R_1)$ and $(s_1, s_2)$ parameter spaces. The field distributions of such modes indicated stronger light confinement in both the in-plane and out-of-plane directions compared with the previous design that did not use $s_2$. In the experimental demonstration, the Si H1 PCN samples exhibited a systematic change in their resonant wavelengths when varying the radial shift of the innermost holes $s_1$ in steps of 1 nm. Their maximum loaded $Q$ factor was $Q_{\rm exp} = 1.2 \times 10^6$, and the corresponding cavity's intrinsic $Q$ factor was $Q_{\rm i} = 1.5 \times 10^6$. Repeating an automated optimization with $(R_0, R_1, s_1)$ for different values of the radial shift of the second innermost holes $s_2$ resulted in $Q_{\rm th} = 4.5 \times 10^8$, a more than a hundred-fold improvement compared with the previous studies. We also discussed some of the major elements that degrade $Q_{\rm exp}$ in reality and estimated the order of practically obtainable $Q_{\rm exp}$. Our work spotlights the power of modal symmetry for improving the performance of nanocavities. It also shows the potential of the H1 PCN in various applications such as functional photonic devices, quantum information processing, and large-scale one- and two-dimensional resonator lattices for studying non-Hermitian and topological photonics and other emergent topics.

\begin{backmatter}
\bmsection{Funding} JSPS KAKENHI Grant Number JP20H05641.

\bmsection{Acknowledgements} We thank Toshiaki Tamamura, Junichi Asaoka, Osamu Moriwaki, Toshifumi Watanabe and Mizuki Ikeya for support with the sample fabrication. We are also grateful to Hideaki Taniyama for support with the complemental FDTD simulation and Shota Kita for fruitful discussion.

\bmsection{Disclosures} The authors declare no conflicts of interest.

\bmsection{Data availability} Data underlying the results presented in this paper are not publicly available at this time but may be obtained from the authors upon reasonable request.
\end{backmatter}



\begin{thebibliography}{10}
	\newcommand{\enquote}[1]{``#1''}
	
	\bibitem{PhC2008}
	J.~D. Joannopoulos, S.~G. Johnson, J.~N. Winn, and R.~D. Meade, \emph{Photonic
		Crystals: Molding the Flow of Light} (Princeton University Press, Princeton,
	2008), 2nd ed.
	
	\bibitem{Painter1999}
	O.~Painter, R.~K. Lee, A.~Scherer, A.~Yariv, J.~D. O'Brien, P.~D. Dapkus, and
	I.~Kim, \enquote{Two-dimensional photonic band-gap defect mode laser,}
	{\protect\JournalTitle{Science}} \textbf{284}, 1819--1821 (1999).
	
	\bibitem{Srinivasan2002}
	K.~Srinivasan and O.~Painter, \enquote{Momentum space design of high-{Q}
		photonic crystal optical cavities,} {\protect\JournalTitle{Opt. Express}}
	\textbf{10}, 670--684 (2002).
	
	\bibitem{Akahane2003}
	Y.~Akahane, T.~Asano, B.-S. Song, and S.~Noda, \enquote{High-{Q} photonic
		nanocavity in a two-dimensional photonic crystal,}
	{\protect\JournalTitle{Nature}} \textbf{425}, 944--947 (2003).
	
	\bibitem{Notomi2004}
	M.~Notomi, A.~Shinya, S.~Mitsugi, E.~Kuramochi, and H.-Y. Ryu,
	\enquote{Waveguides, resonators and their coupled elements in photonic
		crystal slabs,} {\protect\JournalTitle{Opt. Express}} \textbf{12}, 1551--1561
	(2004).
	
	\bibitem{Yoshie2004}
	T.~Yoshie, A.~Scherer, J.~Hendrickson, G.~Khitrova, H.~M. Gibbs, G.~Rupper,
	C.~Ell, O.~B. Shchekin, and D.~G. Deppe, \enquote{Vacuum {R}abi splitting
		with a single quantum dot in a photonic crystal nanocavity,}
	{\protect\JournalTitle{Nature}} \textbf{432}, 200--203 (2004).
	
	\bibitem{Song2005}
	B.-S. Song, S.~Noda, T.~Asano, and Y.~Akahane, \enquote{Ultra-high-{Q} photonic
		double-heterostructure nanocavity,} {\protect\JournalTitle{Nat. Mater.}}
	\textbf{4}, 207--210 (2005).
	
	\bibitem{Englund2005}
	D.~Englund, I.~Fushman, and J.~Vuckovic, \enquote{General recipe for designing
		photonic crystal cavities,} {\protect\JournalTitle{Opt. Express}}
	\textbf{13}, 5961--5975 (2005).
	
	\bibitem{Kuramochi2006}
	E.~Kuramochi, M.~Notomi, S.~Mitsugi, A.~Shinya, T.~Tanabe, and T.~Watanabe,
	\enquote{Ultrahigh-{Q} photonic crystal nanocavities realized by the local
		width modulation of a line defect,} {\protect\JournalTitle{Applied Physics
			Letters}} \textbf{88}, 041112 (2006).
	
	\bibitem{Takahashi2007}
	Y.~Takahashi, H.~Hagino, Y.~Tanaka, B.-S. Song, T.~Asano, and S.~Noda,
	\enquote{High-{Q} nanocavity with a 2-ns photon lifetime,}
	{\protect\JournalTitle{Opt. Express}} \textbf{15}, 17206--17213 (2007).
	
	\bibitem{Kuramochi2008}
	E.~Kuramochi, H.~Taniyama, T.~Tanabe, A.~Shinya, and M.~Notomi,
	\enquote{Ultrahigh-{Q} two-dimensional photonic crystal slab nanocavities in
		very thin barriers,} {\protect\JournalTitle{Applied Physics Letters}}
	\textbf{93}, 111112 (2008).
	
	\bibitem{Notomi2008}
	M.~Notomi, E.~Kuramochi, and H.~Taniyama, \enquote{Ultrahigh-{Q} nanocavity
		with {1D} photonic gap,} {\protect\JournalTitle{Opt. Express}} \textbf{16},
	11095--11102 (2008).
	
	\bibitem{Matsuo2010}
	S.~Matsuo, A.~Shinya, T.~Kakitsuka, K.~Nozaki, T.~Segawa, T.~Sato,
	Y.~Kawaguchi, and M.~Notomi, \enquote{High-speed ultracompact buried
		heterostructure photonic-crystal laser with 13 {fJ} of energy consumed per
		bit transmitted,} {\protect\JournalTitle{Nature Photonics}} \textbf{4},
	648--654 (2010).
	
	\bibitem{Takeda2013}
	K.~Takeda, T.~Sato, A.~Shinya, K.~Nozaki, W.~Kobayashi, H.~Taniyama, M.~Notomi,
	K.~Hasebe, T.~Kakitsuka, and S.~Matsuo, \enquote{Few-{fJ}/bit data
		transmissions using directly modulated lambda-scale embedded active region
		photonic-crystal lasers,} {\protect\JournalTitle{Nature Photonics}}
	\textbf{7}, 569--575 (2013).
	
	\bibitem{Shakoor2014}
	A.~Shakoor, K.~Nozaki, E.~Kuramochi, K.~Nishiguchi, A.~Shinya, and M.~Notomi,
	\enquote{Compact {1D}-silicon photonic crystal electro-optic modulator
		operating with ultra-low switching voltage and energy,}
	{\protect\JournalTitle{Opt. Express}} \textbf{22}, 28623--28634 (2014).
	
	\bibitem{Notomi2005}
	M.~Notomi, A.~Shinya, S.~Mitsugi, G.~Kira, E.~Kuramochi, and T.~Tanabe,
	\enquote{Optical bistable switching action of si high-{Q} photonic-crystal
		nanocavities,} {\protect\JournalTitle{Opt. Express}} \textbf{13}, 2678--2687
	(2005).
	
	\bibitem{Matsuda2011}
	N.~Matsuda, T.~Kato, K.~ichi Harada, H.~Takesue, E.~Kuramochi, H.~Taniyama, and
	M.~Notomi, \enquote{Slow light enhanced optical nonlinearity in a silicon
		photonic crystal coupled-resonator optical waveguide,}
	{\protect\JournalTitle{Opt. Express}} \textbf{19}, 19861--19874 (2011).
	
	\bibitem{Takahashi2013}
	Y.~Takahashi, Y.~Inui, M.~Chihara, T.~Asano, R.~Terawaki, and S.~Noda,
	\enquote{A micrometre-scale {R}aman silicon laser with a microwatt
		threshold,} {\protect\JournalTitle{Nature}} \textbf{498}, 470--474 (2013).
	
	\bibitem{Englund2005_2}
	D.~Englund, D.~Fattal, E.~Waks, G.~Solomon, B.~Zhang, T.~Nakaoka, Y.~Arakawa,
	Y.~Yamamoto, and J.~Vu\ifmmode \check{c}\else
	\v{c}\fi{}kovi\ifmmode~\acute{c}\else \'{c}\fi{}, \enquote{Controlling the
		spontaneous emission rate of single quantum dots in a two-dimensional
		photonic crystal,} {\protect\JournalTitle{Phys. Rev. Lett.}} \textbf{95},
	013904 (2005).
	
	\bibitem{Nomura2010}
	M.~Nomura, N.~Kumagai, S.~Iwamoto, Y.~Ota, and Y.~Arakawa, \enquote{Laser
		oscillation in a strongly coupled single-quantum-dot–nanocavity system,}
	{\protect\JournalTitle{Nature Physics}} \textbf{6}, 279--283 (2010).
	
	\bibitem{Liu2018}
	F.~Liu, A.~J. Brash, J.~O’Hara, L.~M. P.~P. Martins, C.~L. Phillips, R.~J.
	Coles, B.~Royall, E.~Clarke, C.~Bentham, N.~Prtljaga, I.~E. Itskevich, L.~R.
	Wilson, M.~S. Skolnick, and A.~M. Fox, \enquote{High {P}urcell factor
		generation of indistinguishable on-chip single photons,}
	{\protect\JournalTitle{Nature Nanotechnology}} \textbf{13}, 835--840 (2018).
	
	\bibitem{Yariv1999}
	A.~Yariv, Y.~Xu, R.~K. Lee, and A.~Scherer, \enquote{Coupled-resonator optical
		waveguide: a proposal and analysis,} {\protect\JournalTitle{Opt. Lett.}}
	\textbf{24}, 711--713 (1999).
	
	\bibitem{Notomi2008_2}
	M.~Notomi, E.~Kuramochi, and T.~Tanabe, \enquote{Large-scale arrays of
		ultrahigh-{Q} coupled nanocavities,} {\protect\JournalTitle{Nature
			Photonics}} \textbf{2}, 741--747 (2008).
	
	\bibitem{Kuramochi2018}
	E.~Kuramochi, N.~Matsuda, K.~Nozaki, A.~H.~K. Park, H.~Takesue, and M.~Notomi,
	\enquote{Wideband slow short-pulse propagation in one-thousand slantingly
		coupled {L3} photonic crystal nanocavities,} {\protect\JournalTitle{Opt.
			Express}} \textbf{26}, 9552--9564 (2018).
	
	\bibitem{Tanabe2005}
	T.~Tanabe, M.~Notomi, S.~Mitsugi, A.~Shinya, and E.~Kuramochi,
	\enquote{All-optical switches on a silicon chip realized using photonic
		crystal nanocavities,} {\protect\JournalTitle{Applied Physics Letters}}
	\textbf{87}, 151112 (2005).
	
	\bibitem{Nozaki2010}
	K.~Nozaki, T.~Tanabe, A.~Shinya, S.~Matsuo, T.~Sato, H.~Taniyama, and
	M.~Notomi, \enquote{Sub-femtojoule all-optical switching using a
		photonic-crystal nanocavity,} {\protect\JournalTitle{Nature Photonics}}
	\textbf{4}, 477--483 (2010).
	
	\bibitem{Nozaki2013}
	K.~Nozaki, A.~Shinya, S.~Matsuo, T.~Sato, E.~Kuramochi, and M.~Notomi,
	\enquote{Ultralow-energy and high-contrast all-optical switch involving
		{F}ano resonance based on coupled photonic crystal nanocavities,}
	{\protect\JournalTitle{Opt. Express}} \textbf{21}, 11877--11888 (2013).
	
	\bibitem{Tanabe2007}
	T.~Tanabe, M.~Notomi, E.~Kuramochi, A.~Shinya, and H.~Taniyama,
	\enquote{Trapping and delaying photons for one nanosecond in an ultrasmall
		high-{Q} photonic-crystal nanocavity,} {\protect\JournalTitle{Nature
			Photonics}} \textbf{1}, 49--52 (2006).
	
	\bibitem{Nozaki2012}
	K.~Nozaki, A.~Shinya, S.~Matsuo, Y.~Suzaki, T.~Segawa, T.~Sato, Y.~Kawaguchi,
	R.~Takahashi, and M.~Notomi, \enquote{Ultralow-power all-optical {RAM} based
		on nanocavities,} {\protect\JournalTitle{Nature Photonics}} \textbf{6},
	248--252 (2012).
	
	\bibitem{Kuramochi2014}
	E.~Kuramochi, K.~Nozaki, A.~Shinya, K.~Takeda, T.~Sato, S.~Matsuo, H.~Taniyama,
	H.~Sumikura, and M.~Notomi, \enquote{Large-scale integration of
		wavelength-addressable all-optical memories on a photonic crystal chip,}
	{\protect\JournalTitle{Nature Photonics}} \textbf{8}, 474--481 (2014).
	
	\bibitem{Nozaki2019}
	K.~Nozaki, S.~Matsuo, T.~Fujii, K.~Takeda, A.~Shinya, E.~Kuramochi, and
	M.~Notomi, \enquote{Femtofarad optoelectronic integration demonstrating
		energy-saving signal conversion and nonlinear functions,}
	{\protect\JournalTitle{Nature Photonics}} \textbf{13}, 454--459 (2019).
	
	\bibitem{Ryu2003}
	H.-Y. Ryu, M.~Notomi, and Y.-H. Lee, \enquote{High-quality-factor and
		small-mode-volume hexapole modes in photonic-crystal-slab nanocavities,}
	{\protect\JournalTitle{Applied Physics Letters}} \textbf{83}, 4294--4296
	(2003).
	
	\bibitem{Kim2004}
	G.-H. Kim, Y.-H. Lee, A.~Shinya, and M.~Notomi, \enquote{Coupling of small,
		low-loss hexapole mode with photonic crystal slab waveguide mode,}
	{\protect\JournalTitle{Opt. Express}} \textbf{12}, 6624--6631 (2004).
	
	\bibitem{Tanabe2007_2}
	T.~Tanabe, A.~Shinya, E.~Kuramochi, S.~Kondo, H.~Taniyama, and M.~Notomi,
	\enquote{Single point defect photonic crystal nanocavity with ultrahigh
		quality factor achieved by using hexapole mode,}
	{\protect\JournalTitle{Applied Physics Letters}} \textbf{91}, 021110 (2007).
	
	\bibitem{Takagi2012}
	H.~Takagi, Y.~Ota, N.~Kumagai, S.~Ishida, S.~Iwamoto, and Y.~Arakawa,
	\enquote{High-{Q} {H1} photonic crystal nanocavities with efficient vertical
		emission,} {\protect\JournalTitle{Opt. Express}} \textbf{20}, 28292--28300
	(2012).
	
	\bibitem{Sakoda2005}
	K.~Sakoda, \emph{Optical Properties of Photonic Crystals}, Springer Series in
	Optical Sciences (Springer-Verlag, Berlin, Heidelberg, 2005), 2nd ed.
	
	\bibitem{Altug2004}
	H.~Altug and J.~Vučković, \enquote{Two-dimensional coupled photonic crystal
		resonator arrays,} {\protect\JournalTitle{Applied Physics Letters}}
	\textbf{84}, 161--163 (2004).
	
	\bibitem{Takata2017}
	K.~Takata and M.~Notomi, \enquote{{PT}-symmetric coupled-resonator waveguide
		based on buried heterostructure nanocavities,} {\protect\JournalTitle{Phys.
			Rev. Applied}} \textbf{7}, 054023 (2017).
	
	\bibitem{Takata2018}
	K.~Takata and M.~Notomi, \enquote{Photonic topological insulating phase induced
		solely by gain and loss,} {\protect\JournalTitle{Phys. Rev. Lett.}}
	\textbf{121}, 213902 (2018).
	
	\bibitem{Han2019}
	C.~Han, M.~Lee, S.~Callard, C.~Seassal, and H.~Jeon, \enquote{Lasing at
		topological edge states in a photonic crystal {L3} nanocavity dimer array,}
	{\protect\JournalTitle{Light: Sci. Appl.}} \textbf{8}, 40 (2019).
	
	\bibitem{Duggan2020}
	R.~Duggan, S.~A. Mann, and A.~Al\`u, \enquote{Nonreciprocal photonic
		topological order driven by uniform optical pumping,}
	{\protect\JournalTitle{Phys. Rev. B}} \textbf{102}, 100303 (2020).
	
	\bibitem{Takata2021}
	K.~Takata, K.~Nozaki, E.~Kuramochi, S.~Matsuo, K.~Takeda, T.~Fujii, S.~Kita,
	A.~Shinya, and M.~Notomi, \enquote{Observing exceptional point degeneracy of
		radiation with electrically pumped photonic crystal coupled-nanocavity
		lasers,} {\protect\JournalTitle{Optica}} \textbf{8}, 184--192 (2021).
	
	\bibitem{Fong2021}
	C.~F. Fong, Y.~Ota, Y.~Arakawa, S.~Iwamoto, and Y.~K. Kato, \enquote{Chiral
		modes near exceptional points in symmetry broken {H1} photonic crystal
		cavities,} {\protect\JournalTitle{Phys. Rev. Research}} \textbf{3}, 043096
	(2021).
	
	\bibitem{Takata2022}
	K.~Takata, N.~Roberts, A.~Shinya, and M.~Notomi, \enquote{Imaginary couplings
		in non-{H}ermitian coupled-mode theory: {E}ffects on exceptional points of
		optical resonators,} {\protect\JournalTitle{Phys. Rev. A}} \textbf{105},
	013523 (2022).
	
	\bibitem{Hentinger2022}
	F.~Hentinger, M.~Hedir, B.~Garbin, M.~Marconi, L.~Ge, F.~Raineri, J.~A.
	Levenson, and A.~M. Yacomotti, \enquote{Direct observation of zero modes in a
		non-{H}ermitian optical nanocavity array,} {\protect\JournalTitle{Photon.
			Res.}} \textbf{10}, 574--586 (2022).
	
	\bibitem{Ozdemir2019}
	{\c{S}}.~K. {\"O}zdemir, S.~Rotter, F.~Nori, and L.~Yang, \enquote{Parity--time
		symmetry and exceptional points in photonics,} {\protect\JournalTitle{Nat.
			Mater.}} \textbf{18}, 783--798 (2019).
	
	\bibitem{Ota2020}
	Y.~Ota, K.~Takata, T.~Ozawa, A.~Amo, Z.~Jia, B.~Kante, M.~Notomi, Y.~Arakawa,
	and S.~Iwamoto, \enquote{Active topological photonics,}
	{\protect\JournalTitle{Nanophotonics}} \textbf{9}, 547--567 (2020).
	
	\bibitem{Szameit2011}
	A.~Szameit, M.~C. Rechtsman, O.~Bahat-Treidel, and M.~Segev,
	\enquote{$\mathcal{P}\mathcal{T}$-symmetry in honeycomb photonic lattices,}
	{\protect\JournalTitle{Phys. Rev. A}} \textbf{84}, 021806 (2011).
	
	\bibitem{Kremer2019}
	M.~Kremer, T.~Biesenthal, L.~J. Maczewsky, M.~Heinrich, R.~Thomale, and
	A.~Szameit, \enquote{Demonstration of a two-dimensional
		$\mathcal{P}\mathcal{T}$-symmetric crystal,} {\protect\JournalTitle{Nature
			Communications}} \textbf{10}, 435 (2019).
	
	\bibitem{Wu2015}
	L.-H. Wu and X.~Hu, \enquote{Scheme for achieving a topological photonic
		crystal by using dielectric material,} {\protect\JournalTitle{Phys. Rev.
			Lett.}} \textbf{114}, 223901 (2015).
	
	\bibitem{Noh2018}
	J.~Noh, W.~A. Benalcazar, S.~Huang, M.~J. Collins, K.~P. Chen, T.~L. Hughes,
	and M.~C. Rechtsman, \enquote{Topological protection of photonic mid-gap
		defect modes,} {\protect\JournalTitle{Nature Photonics}} \textbf{12},
	408--415 (2018).
	
	\bibitem{Li2020}
	M.~Li, D.~Zhirihin, M.~Gorlach, X.~Ni, D.~Filonov, A.~Slobozhanyuk, A.~Alù,
	and A.~B. Khanikaev, \enquote{Higher-order topological states in photonic
		kagome crystals with long-range interactions,} {\protect\JournalTitle{Nature
			Photonics}} \textbf{14}, 89--94 (2020).
	
	\bibitem{Khanikaev2017}
	A.~B. Khanikaev and G.~Shvets, \enquote{Two-dimensional topological photonics,}
	{\protect\JournalTitle{Nature Photonics}} \textbf{11}, 763--773 (2017).
	
	\bibitem{Minkov2014}
	M.~Minkov and V.~Savona, \enquote{Automated optimization of photonic crystal
		slab cavities,} {\protect\JournalTitle{Sci. Rep.}} \textbf{4}, 5124 (2014).
	
	\bibitem{Taguchi2011}
	Y.~Taguchi, Y.~Takahashi, Y.~Sato, T.~Asano, and S.~Noda, \enquote{Statistical
		studies of photonic heterostructure nanocavities with an average {Q} factor
		of three million,} {\protect\JournalTitle{Opt. Express}} \textbf{19},
	11916--11921 (2011).
	
	\bibitem{Lai2014}
	Y.~Lai, S.~Pirotta, G.~Urbinati, D.~Gerace, M.~Minkov, V.~Savona, A.~Badolato,
	and M.~Galli, \enquote{Genetically designed {L3} photonic crystal
		nanocavities with measured quality factor exceeding one million,}
	{\protect\JournalTitle{Applied Physics Letters}} \textbf{104}, 241101 (2014).
	
	\bibitem{Simbula2017}
	A.~Simbula, M.~Schatzl, L.~Zagaglia, F.~Alpeggiani, L.~C. Andreani,
	F.~Schäffler, T.~Fromherz, M.~Galli, and D.~Gerace, \enquote{Realization of
		high-{Q/V} photonic crystal cavities defined by an effective
		{A}ubry-{A}ndré-{H}arper bichromatic potential,} {\protect\JournalTitle{APL
			Photonics}} \textbf{2}, 056102 (2017).
	
	\bibitem{Benevides2017}
	R.~Benevides, F.~G.~S. Santos, G.~O. Luiz, G.~S. Wiederhecker, and T.~P.~M.
	Alegre, \enquote{Ultrahigh-{Q} optomechanical crystal cavities fabricated in
		a {CMOS} foundry,} {\protect\JournalTitle{Sci. Rep.}} \textbf{7}, 2491
	(2017).
	
	\bibitem{Ashida2018}
	K.~Ashida, M.~Okano, T.~Yasuda, M.~Ohtsuka, M.~Seki, N.~Yokoyama, K.~Koshino,
	K.~Yamada, and Y.~Takahashi, \enquote{Photonic crystal nanocavities with an
		average {Q} factor of 1.9 million fabricated on a 300-mm-wide {SOI} wafer
		using a {CMOS}-compatible process,} {\protect\JournalTitle{J. Lightwave
			Technol.}} \textbf{36}, 4774--4782 (2018).
	
	\bibitem{comsol}
	\enquote{{COMSOL Multiphysics\textregistered},} \url{https://www.comsol.com/}.
	
	\bibitem{Johnson2001}
	S.~G. Johnson, S.~Fan, A.~Mekis, and J.~D. Joannopoulos,
	\enquote{Multipole-cancellation mechanism for high-{Q} cavities in the
		absence of a complete photonic band gap,} {\protect\JournalTitle{Applied
			Physics Letters}} \textbf{78}, 3388--3390 (2001).
	
	\bibitem{Minkov2017}
	M.~Minkov, V.~Savona, and D.~Gerace, \enquote{Photonic crystal slab cavity
		simultaneously optimized for ultra-high {Q/V} and vertical radiation
		coupling,} {\protect\JournalTitle{Applied Physics Letters}} \textbf{111},
	131104 (2017).
	
	\bibitem{Asano2018}
	T.~Asano and S.~Noda, \enquote{Optimization of photonic crystal nanocavities
		based on deep learning,} {\protect\JournalTitle{Opt. Express}} \textbf{26},
	32704--32717 (2018).
	
	\bibitem{Shibata2021}
	T.~Shibata, T.~Asano, and S.~Noda, \enquote{Fabrication and characterization of
		an {L3} nanocavity designed by an iterative machine-learning method,}
	{\protect\JournalTitle{APL Photonics}} \textbf{6}, 036113 (2021).
	
	\bibitem{Vasco2021}
	J.~P. Vasco and V.~Savona, \enquote{Global optimization of an encapsulated
		{Si}/{SiO}$_2$ {L3} cavity with a 43 million quality factor,}
	{\protect\JournalTitle{Sci. Rep.}} \textbf{11}, 10121 (2021).
	
	\bibitem{Tanaka2008}
	Y.~Tanaka, T.~Asano, and S.~Noda, \enquote{Design of photonic crystal
		nanocavity with {Q}-factor of $\sim 10^9$,} {\protect\JournalTitle{J.
			Lightwave Technol.}} \textbf{26}, 1532--1539 (2008).
	
	\bibitem{Nakamura2016}
	T.~Nakamura, Y.~Takahashi, Y.~Tanaka, T.~Asano, and S.~Noda,
	\enquote{Improvement in the quality factors for photonic crystal nanocavities
		via visualization of the leaky components,} {\protect\JournalTitle{Opt.
			Express}} \textbf{24}, 9541--9549 (2016).
	
	\bibitem{Dharanipathy2014}
	U.~P. Dharanipathy, M.~Minkov, M.~Tonin, V.~Savona, and R.~Houdré,
	\enquote{High-q silicon photonic crystal cavity for enhanced optical
		nonlinearities,} {\protect\JournalTitle{Applied Physics Letters}}
	\textbf{105}, 101101 (2014).
	
	\bibitem{Hagino2009}
	H.~Hagino, Y.~Takahashi, Y.~Tanaka, T.~Asano, and S.~Noda, \enquote{Effects of
		fluctuation in air hole radii and positions on optical characteristics in
		photonic crystal heterostructure nanocavities,} {\protect\JournalTitle{Phys.
			Rev. B}} \textbf{79}, 085112 (2009).
	
	\bibitem{Hughes2005}
	S.~Hughes, L.~Ramunno, J.~F. Young, and J.~E. Sipe, \enquote{Extrinsic optical
		scattering loss in photonic crystal waveguides: Role of fabrication disorder
		and photon group velocity,} {\protect\JournalTitle{Phys. Rev. Lett.}}
	\textbf{94}, 033903 (2005).
	
	\bibitem{Nelder1965}
	J.~A. Nelder and R.~Mead, \enquote{{A Simplex Method for Function
			Minimization},} {\protect\JournalTitle{The Computer Journal}} \textbf{7},
	308--313 (1965).
	
	\bibitem{Asano2017}
	T.~Asano, Y.~Ochi, Y.~Takahashi, K.~Kishimoto, and S.~Noda, \enquote{Photonic
		crystal nanocavity with a {Q} factor exceeding eleven million,}
	{\protect\JournalTitle{Opt. Express}} \textbf{25}, 1769--1777 (2017).
	
	\bibitem{Yamashita1996}
	Y.~Yamashita, K.~Namba, Y.~Nakato, Y.~Nishioka, and H.~Kobayashi,
	\enquote{Spectroscopic observation of interface states of ultrathin silicon
		oxide,} {\protect\JournalTitle{Journal of Applied Physics}} \textbf{79},
	7051--7057 (1996).
	
	\bibitem{Borselli2006}
	M.~Borselli, T.~J. Johnson, and O.~Painter, \enquote{Measuring the role of
		surface chemistry in silicon microphotonics,} {\protect\JournalTitle{Applied
			Physics Letters}} \textbf{88}, 131114 (2006).
	
	\bibitem{Yablonovitch1986}
	E.~Yablonovitch, D.~L. Allara, C.~C. Chang, T.~Gmitter, and T.~B. Bright,
	\enquote{Unusually low surface-recombination velocity on silicon and
		germanium surfaces,} {\protect\JournalTitle{Phys. Rev. Lett.}} \textbf{57},
	249--252 (1986).
	
	\bibitem{Takahagi1988}
	T.~Takahagi, I.~Nagai, A.~Ishitani, H.~Kuroda, and Y.~Nagasawa, \enquote{The
		formation of hydrogen passivated silicon single‐crystal surfaces using
		ultraviolet cleaning and {HF} etching,} {\protect\JournalTitle{Journal of
			Applied Physics}} \textbf{64}, 3516--3521 (1988).
	
	\bibitem{Sekoguchi2014}
	H.~Sekoguchi, Y.~Takahashi, T.~Asano, and S.~Noda, \enquote{Photonic crystal
		nanocavity with a {Q}-factor of ~9 million,} {\protect\JournalTitle{Opt.
			Express}} \textbf{22}, 916--924 (2014).
	
	\bibitem{Lee2001}
	K.~K. Lee, D.~R. Lim, L.~C. Kimerling, J.~Shin, and F.~Cerrina,
	\enquote{Fabrication of ultralow-loss {Si}/{SiO}$_2$ waveguides by roughness
		reduction,} {\protect\JournalTitle{Opt. Lett.}} \textbf{26}, 1888--1890
	(2001).
	
	\bibitem{Sparacin2005}
	D.~K. Sparacin, S.~J. Spector, and L.~C. Kimerling, \enquote{Silicon waveguide
		sidewall smoothing by wet chemical oxidation,} {\protect\JournalTitle{J.
			Lightwave Technol.}} \textbf{23}, 2455 (2005).
	
	\bibitem{Ling1992}
	L.~Ling, Z.~J. Radzimski, T.~Abe, and F.~Shimura, \enquote{The effect of bonded
		interface on electrical properties of bonded silicon‐on‐insulator
		wafers,} {\protect\JournalTitle{Journal of Applied Physics}} \textbf{72},
	3610--3616 (1992).
	
	\bibitem{Panuski2020}
	C.~Panuski, D.~Englund, and R.~Hamerly, \enquote{Fundamental thermal noise
		limits for optical microcavities,} {\protect\JournalTitle{Phys. Rev. X}}
	\textbf{10}, 041046 (2020).
	
	\bibitem{Tanabe2007_3}
	T.~Tanabe, M.~Notomi, E.~Kuramochi, and H.~Taniyama, \enquote{Large pulse delay
		and small group velocity achieved using ultrahigh-{Q} photonic crystal
		nanocavities,} {\protect\JournalTitle{Opt. Express}} \textbf{15}, 7826--7839
	(2007).
	
\end{thebibliography}






\end{document}